\documentclass[twoside]{article}     % onecolumn (ditto)
\usepackage{cite}
\usepackage{graphicx}
\usepackage[a4paper]{geometry}
\usepackage[latin1]{inputenc}
\usepackage[T1]{fontenc}
\usepackage{lmodern}
\usepackage{RR}
\usepackage{float}
\usepackage{microtype}
\usepackage{subfig}
\usepackage{booktabs}
\usepackage{tikz}
\usepackage{url}
\usepackage{color}
\usepackage{ulem}
\usepackage{mathtools}
\usepackage{soul}
\usepackage{xspace}
\usepackage{algorithm}
\usepackage{algorithmic}
\usepackage{comment}

\RRNo{9325}
\RRdate{Octobre 2020}

\RRauthor{% les auteurs
Antoine Boutet\thanks[inria]{Univ Lyon, INSA Lyon, Inria, CITI, F-69621 VILLEURBANNE, France} %
\and Carole Frindel\thanks[creatis]{Univ Lyon, INSA Lyon, CNRS, Inserm, CREATIS UMR 5220, U1206, F-69621 VILLEURBANNE, France} %
\and S\'ebastien Gambs\thanks[uqam]{Universit\'e du Qu\'ebec \`a Montr\'eal, Montr\'eal, Qu\'ebec, Canada} 
\and Th\'eo Jourdan\thanksref{inria}\thanksref{creatis} 
\and Rosin Claude Ngueveu\thanksref{uqam}
}

\authorhead{Boutet \& Frindel \& others}
\RRtitle{\name: Assainissement dynamique des donn\'ees de capteur de mouvement contre l'inf\'erence d'informations sensibles \`a partir de r\'eseaux adversariaux}
\RRetitle{\name: Dynamically sanitizing motion sensor data against sensitive inferences through adversarial networks}

\titlehead{\name}

\RRresume{Avec l'adoption g\'en\'eralis\'ee du suivi d'activit\'e, un nombre croissant d'utilisateurs s'appuient sur des applications mobiles pour surveiller leur activit\'e physique par le biais de leur smartphone. 
Le fait d'accorder aux applications un acc\`es direct aux donn\'ees des capteurs expose les utilisateurs \`a des risques pour leur vie priv\'ee.
En effet, ces donn\'ees de capteurs de mouvement sont g\'en\'eralement transmises \`a des applications d'analyse h\'eberg\'ees sur le cloud, qui exploitent des mod\`eles d'apprentissage machine pour fournir aux utilisateurs un retour d'information sur leur sant\'e. 
Cependant, rien n'emp\^eche le fournisseur de services d'inf\'erer des informations priv\'ees et potentiellement sensibles sur un utilisateur, telles que des attributs d\'emographiques ou de sant\'e.

Dans cet article, nous pr\'esentons \name, un syst\`eme de pr\'eservation de la vie priv\'ee pour assainir les donn\'ees provenant de capteurs de mouvement contre les inf\'erence non d\'esir\'ees d'informations sensibles (c'est-\`a-dire am\'eliorer la vie priv\'ee) tout en limitant la perte de pr\'ecision sur la surveillance de l'activit\'e physique (c'est-\`a-dire maintenir une certaine utilit\'e dans les donn\'ees prot\'eg\'ees).
Pour garantir un bon compromis entre utilit\'e et respect de la vie priv\'ee, \name s'appuie sur des R\'eseaux g\'en\'eratifs Adversariaux (GAN) pour assainir les donn\'ees issues des capteurs. 
Plus pr\'ecis\'ement, en apprenant de mani\`ere comp\'etitive plusieurs r\'eseaux, \name est capable de construire des mod\`eles d'apprentissage machine qui assainissent les donn\'ees de mouvement contre l'inf\'erence d'un attribut sensible sp\'ecifi\'e (par exemple, le genre) tout en maintenant une grande pr\'ecision sur la reconnaissance d'activit\'e.
De plus, \name construit divers mod\`eles d'assainissement, caract\'eris\'es par diff\'erents ensembles d'hyperparam\`etres dans la fonction de perte globale, pour proposer un sch\'ema d'apprentissage du transfert dans le temps en s\'electionnant dynamiquement le mod\`ele qui offre le meilleur compromis entre utilit\'e et respect de la vie priv\'ee en fonction des donn\'ees entrantes.

Les exp\'eriences men\'ees sur des ensembles de donn\'ees r\'eels montrent que \name peut limiter consid\'erablement l'inf\'erence de genre jusqu'\`a 41\% (de 98\% avec des donn\'ees brutes \`a 57\% avec des donn\'ees assainies) tout en ne r\'eduisant la pr\'ecision de la reconnaissance d'activit\'e que de 3\% (de 95\% avec des donn\'ees brutes \`a 92\% avec des donn\'ees assainies)
}
\RRabstract{
With the widespread adoption of the quantified self movement, an increasing number of users rely on mobile applications to monitor their physical activity through their smartphones. 
However, granting applications a direct access to sensor data expose users to privacy risks.
In particular, motion sensor data are usually transmitted to analytics applications hosted on the cloud, which leverages on machine learning models to provide feedback on their activity status to users.
In this setting, nothing prevents the service provider to infer private and sensitive information about a user such as health or demographic attributes.
To address this issue, we propose \name, a privacy-preserving framework to sanitize motion sensor data against unwanted sensitive inferences (\textit{i.e.}, improving privacy) while limiting the loss of accuracy on the physical activity monitoring (\textit{i.e.}, maintaining data utility).
Our approach is inspired from the framework of Generative Adversarial Networks to sanitize the sensor data for the purpose of ensuring a good trade-off between utility and privacy.
More precisely, by learning in a competitive manner several networks, \name is able to build models that sanitize motion data against inferences on a specified sensitive attribute (\textit{e.g.}, gender) while maintaining an accurate activity recognition.
\name builds various sanitizing models, characterized by different sets of hyperparameters in the global loss function, to propose a transfer learning scheme over time by dynamically selecting the model which provides the best utility and privacy trade-off according to the incoming data.
Experiments conducted on real datasets demonstrate that \name can drastically limit the gender inference up to 41\% (from 98\% with raw data to 57\% with sanitized data) while only reducing the accuracy of activity recognition by 3\% (from 95\% with raw data to 92\% with sanitized data).
}
\RRmotcle{vie priv\'ee, intelligence artificielle, transparence, donn\'ees de sant\'e, confidentialit\'e}

\RRkeyword{privacy, articifial intelligence, transparency, health data, confidentiality}
\RRprojets{Privatics}
\URRhoneAlpes % pour ceux qui sont dans les montagnes

\begin{document}

%\runningtitle{DYSAN}

\newcommand{\name}{\textsc{DySan}\xspace}
\newcommand{\myBer}{$BER$}
\newcommand{\mySac}{sAcc}
\newcommand{\sandist}{fid}
\newcommand{\mdist}{div}
\newcommand{\tAcc}{tAcc}
\newcommand{\dataset}{\textit{X}}
\newcommand{\sensAttr}{\textit{S}}
\newcommand{\sensAttrL}{\textit{s}}
\newcommand{\decAttr}{\textit{Y}}
\newcommand{\decAttrL}{y}
\newcommand{\notSensAttr}{\textit{A}}
\newcommand{\notSensAttrL}{\textit{a}}
\newcommand{\dis}{$D_{isc}$}
\newcommand{\san}{$S_{an}$}
\newcommand{\utp}{$U_{tp}$}
\newcommand{\utpname}{activity recognizer}
\newcommand{\bodyMeasure}{O}
\newcommand{\ros}[1]{\textcolor{red}{#1}}

\def\rsn#1{\textcolor{cyan}{#1}} % Ajout Rosin
\def\seb#1{\textcolor{red}{#1}} % Ajout Sébastien
\def\teo#1{\textcolor{blue}{#1}}  % Ajout Théo
\def\ant#1{\textcolor{green}{#1}} % Ajout Antoine
\def\crl#1{\textcolor{magenta}{#1}} % Ajout Carole
\def\rque#1{\textcolor{orange}{#1}} % remarque

\makeRR

\section{Introduction}
\label{sec:intro}
    
The integration of motion sensors in smartphones and wearables has been accompanied by the growth of the quantified self movement~\cite{yang2016user}.
%Seb*: si possible je suggère de mettre une référence pour le mouvement quantified-self
%done
For instance nowadays, users increasingly exploit these devices to monitor their physical activities. 
Usually, the motion sensor data are not analyzed directly on the device but are rather transmitted to analytics applications hosted on the cloud. 
These analytics applications leverage machine learning models to compute statistical indicators related to the status of users that are send back to them. 
While these analyses can bring many benefits from the health perspective~\cite{stroke1,stroke2,qi2015survey}, they can also lead to privacy breaches by exposing personal information regarding the individual concerned. 
Indeed, a large range of inferences can be done from motion sensor data including sensitive ones such as demographic and health-related attributes~\cite{han2012accomplice,kroger2019privacy,lee2002activity}.
%Seb*: suggestion: rajouter aussi une référence montrant qu'on peut déduire des informations sur la géolocalisation à partir des données de capteurs
%done

Consider for instance the scenario in which Alice, a woman, uses a fitness application on her smartphone to monitor her physical activity. 
The application performs the activity recognition as well as the activity monitoring on the cloud. 
However even if the service provider declares that it will never do it, Alice has no formal guarantees that her data will not be processed to infer other information about her (\textit{e.g.}, for targeting or marketing purposes). 
Another possible scenario is related to the new trend of insurance companies that propose discount to clients if they accept to use a connected device to follow their daily activity~\cite{tedesco2017review}. 
These data can be used to provide a personalized coaching for better health management but also for early detection of a pathology, which can negatively impact the insurance cost or lead to other type of discrimination.
%The goal of this work is to provide a solution which ensures that the transmitted information can not be used for inferring unwanted sensitive information but only for monitoring the activity of users.
% The goal of this work is to provide a solution which sanitizes 
% the motion sensor data in such a way that it hides sensitive attributes 
% while still preserving the activity information contained in the data.
To address the issues raised by these scenarios, in this work we propose a solution sanitizing the motion sensor data in such a way that it hides sensitive attributes while preserving the activity information contained in the data.

To achieve this objective, we design \name, inspired from the framework of Generative Adversarial Networks (GANs)~\cite{pan2019recent} to sanitize the sensor data.
%Seb*: suggestion mettre une référence pour les GANs
%done
More precisely, by learning in a competitive manner several networks, \name is able to build models sanitizing motion data to prevent inferences on a specified sensitive attribute while maintaining a high level of activity recognition.
%Seb*: ci-dessus je pense qu'on devrait préciser tout de suite pourquoi on a besoin d'entraîné plusieurs réseaux  
In addition, by limiting the distortion between the raw and sanitized data, \name also maintains a high level of utility with respect to other analysis tasks related to activity monitoring (\textit{e.g.}, steps counting). 
%Finally, each time \name processes an incoming batch of data, it is able to dynamically selects the sanitizing model which limits as much as possible the risk of inference of the sensitive attribute.

Furthermore, our approach aims at addressing the heterogeneous aspect of sensor data, which is inherent to the way each user moves, to the characteristics of the device used for data collection and to the evolution of activity during the day.
Thus, as one sanitizing model cannot provide the best utility and privacy trade-off for all users over time, \name builds a set of diverse sanitizing models by exploring different combination of hyperparameters balancing loss functions of activity recognition, sensitive inference and data distortion terms.
By doing so, \name is able to dynamically select the model which provides the best trade-off over time according to the incoming sensor data.

The evaluation of \name on real datasets, in which the \textit{gender} is considered as the sensitive information to hide, demonstrates that \name can drastically limit the gender inference up to $41\%$ while only inducing a drop of $3\%$ on the accuracy of activity recognition.
%Seb*: pour la prédiction du genre, quel serait le taux de base si on prenait la classe majoritaire
In addition to preserve activity recognition, \name, by limiting data distortion, also preserves the sensor data utility for other analytical tasks such as estimating the number of steps. 
Moreover, we show that the dynamic model selection of \name successfully provides an adaptation of the sanitization according to the incoming user data. This dynamic model selection is specially useful to transfer learning from the dataset used to build the sanitizer models to another dataset with new users with potentially different data distribution.
Our dynamic sanitization method overcomes several shortcomings of the state-of-the-art approaches, namely the use of the same sanitization model for all users over time, which may lead to a poor privacy-utility trade-off for atypical users.
Lastly, we evaluate the cost of operating \name on a smartphone and show that the introduced overhead is compatible with real-time processing and that the energy consumption remains reasonable.
Our implementation of \name as well as the datasets used to assess its performances are publicly available~\footnote{\name: \url{https://github.com/DynamicSanitizer/DySan}}

The outline of the paper is as follows. 
First, the problem definition and the considered system model are described in Section~\ref{sec:sysmodel}. Then, \name is presented in Section~\ref{sec:protection} before reviewing the experimental setting as well as the results obtained, respectively in Section~\ref{sec:experiment} and Section~\ref{sec:results}.
Finally, the related work is discussed in Section~\ref{sec:relatedworks} before concluding in Section~\ref{sec:conclusion}.

%citations : \cite{zhang2014iot} (2), \cite{wood:cleartext,DBLP:journals/corr/ArankiB15} (3,4)
% \cite{6956585} (5), \cite{doi:10.1001/jama.285.24.3075} (6), \cite{seref2016opportunities} (7)

\section{Problem definition and system model}
\label{sec:sysmodel}

% ATTENTION nos ne parlons pas d'application de monitoring de patient dans ce papier

%Before presenting the design of \name (Section~\ref{sec:protection}), 
%we describe the considered system model and the problem definition.

We consider a mobile application installed on the user's smartphone aiming to monitor its physical activity.
The smartphone of the user is assumed to be trusted.
For instance, we consider that \name could be deployed in the trusted environment of the smartphone to prevent the mobile application to have a direct access to the sensor data but only from the output of \name (thus ensuring that the mobile application uses only sanitized data).
Afterwards, the mobile application sends the sanitized data to a server hosted on the cloud.
This server leverages machine learning models to identify the activity of the user or to estimate other physical activity features (\textit{e.g.}, number of steps).
The server is considered to follow the honest-but-curious adversary model in the sense that it may also try to infer additional sensitive information from the sensor data.
For the rest of the paper, we consider the \textit{gender} as being the sensitive attribute to protect.
Note however that our approach is much more generic and could be applied to protect other sensitive attributes (\textit{e.g.}, handicap).
This choice is only motivated by the availability of different datasets with this information. Note also that the gender could be inferred from the list of performed activities and their associated frequencies in case of unbalanced data distribution between men and women (which is not the case in the datasets considered in this paper).

We consider raw motion sensor data (denoted by \notSensAttr) captured through accelerometer and gyroscope that sample 3-axial signals with a frequency of 50 Hz.
To enable activity recognition over time, the raw sensor data are split in sliding windows, in which each sliding window is considered to be a sample of a single activity (\textit{i.e.}, by assumption the user cannot perform two different activities during a single sliding window). 
The datasets used are composed of four type of dynamic activities (\textit{i.e.}, walking, running, climbing and going down stairs), and we chose the length of sliding window to match a walking cycle of two steps. 
The choice of the window size is not trivial, especially for an activity recognition task, and has to be well calibrated. 
Indeed, a small window size could split an activity signal while large window size could contain multiple activity signals.
Knowing that in average the walking pace is not less than 1.5 steps per second \cite{steps}, the window length $T$ is chosen to be 2.5 seconds with an overlap of 50 $\%$.

We assume a population of $N$ users contained in a dataset \dataset{} storing all users data. This dataset includes the raw sensor data as well as the label associated to the activity performed by the user (denoted by a multi-valued attribute \decAttr), the binary sensitive attribute (denoted by \sensAttr) and a timestamp.
Thus, the dataset $\dataset{} = \{\notSensAttr{}, \decAttr{}, \sensAttr{}\}$ in which $\notSensAttr{} = (\notSensAttr{}_{1},\ldots, \notSensAttr{}_{T})$.

The objective of \name is to protect the user motion sensor data against sensitive attribute inferences while maintaining data utility.
More formally, we aim at learning a set of sanitizers $S_{an_{\alpha, \lambda, \beta}}$ for various values of the hyperparameters $\alpha$, $\lambda$ and $\beta$.
%Seb* ci-dessus il faudrait expliquer à quoi sert chacun de ces paramètres
Each sanitizer will transform the original data \dataset{} into $\bar{\dataset{}} = S_{an_{\alpha, \lambda, \beta}}(\dataset{}) = \{\bar{\notSensAttr{}},\decAttr{}, \sensAttr{}\}$; $\bar{\notSensAttr{}} = (\bar{\notSensAttr{}}_{1},\bar{\ldots}, \bar{\notSensAttr{}}_{T})$. 
%Seb*
This set of sanitizers is learned so that it is difficult to build a discriminator $D_{isc}$ trained to predict \sensAttr{} from the sanitized data and activities $\{\bar{\notSensAttr{}} \decAttr{}\}$ while an activity predictor $P_{red}$ trained on the same sensor data ($\bar{\notSensAttr{}}$) is able to maintain an accuracy close to the original one. 
To further preserve the utility of $\bar{\dataset{}}$, we also constrain the sanitization process to minimize the distortion between the original and sanitized data.

Furthermore, \name aims to dynamically adapt over time the hyperparameters of the model according to the incoming data of each user. 
Indeed, while a particular model could provide the best utility/privacy trade-off on average for all users with respect to training dataset, the model leading to the best trade-off can change when testing on new user (\textit{e.g.}, when the new user data does not fit the data distribution of the training dataset).
More formally, we aim to find for each window of data the sanitizer $\widehat{S_{an_{\alpha, \lambda, \beta}}}$ providing the best utility/privacy trade-off for the current incoming data. % a considered period which includes a small set of last windows. 
This trade-off is defined by a metric combining the accuracy of the activity recognition and the inference of the sensitive attribute.

\section{DYSAN: Dynamic Sanitizer}
\label{sec:protection}

%In this section, we start by providing an overview of our solution (Section~\ref{sec:overview}) before developing in more details how \name builds multiple sanitizers (Section~\ref{sec:training}) during the training phase (Section~\ref{sec:trainingphase}) and how the dynamic sanitizer model selection is achieved during the online phase (Section~\ref{sec:online}).

%\subsection{Overview}
%\label{sec:overview}

\begin{figure}[h!]
	\centering
	\vspace{-0.8cm}
	\includegraphics[width=9cm]{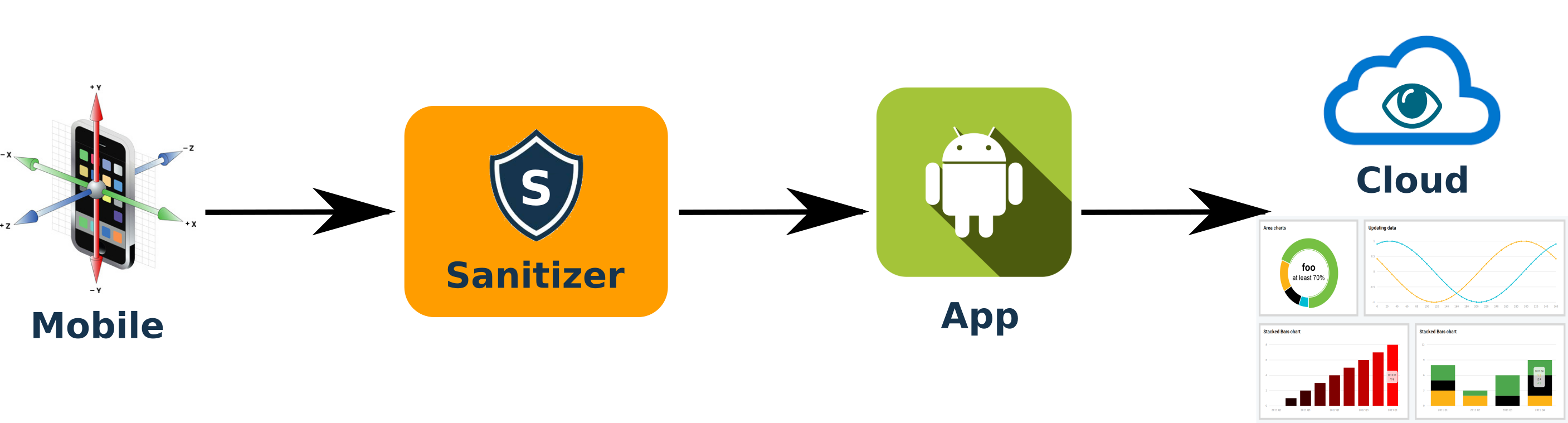}
%	\vspace{-1mm}
	\caption{\small Overview of dynamic sanitizer.}
	\label{fig:chaine}
%	\vspace{-0.3cm}
\end{figure}

\begin{figure*}[t]
	\centering
	\includegraphics[width=14cm]{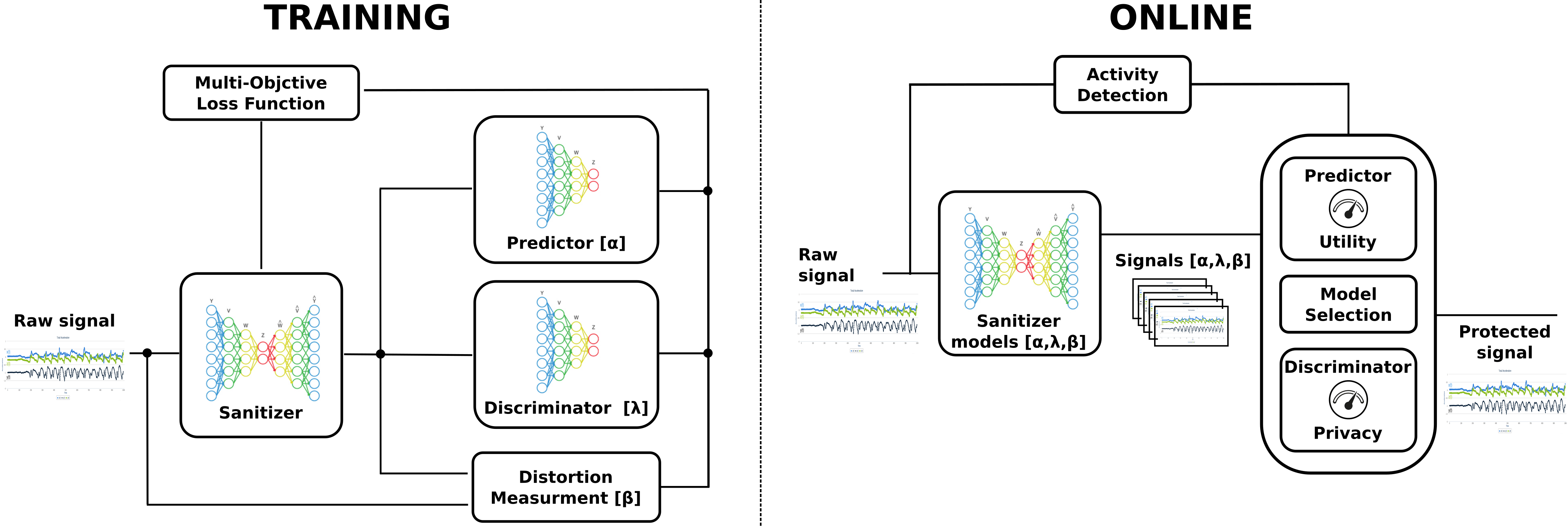}
%	\vspace{-3mm}
	\caption{\small Dynamically sanitizing motion sensor data  with \name framework during the training (left) and the online (right) phases. Training phase allows to build different models that are distinguished by their parameters and online phase allows to choose among these models the most adapted to the final user.}
	\label{fig:archi}
%	\vspace{-0.2cm}
\end{figure*}

An overview of \name is shown in Figure~\ref{fig:chaine}. To avoid an unwanted exploitation of the motion sensor data, these data are sanitized by \name before being transmitted to the mobile application. This sanitizing process removes the correlations with \sensAttr{} in the sensor data while preserving the information necessary to detect the activity performed by a user. In addition, \name also aims at limiting the distortion between the raw and sanitized data to preserve the utility for other analytical tasks.
%\textbf{Feedbacks are provided to users in order to support him in the trade-off between the protection of sensitive information and the activity recognition. TO FIX}
Finally, the resulting sanitized data are sent to an analytics application hosted on the cloud, exploiting machine learning models to classify the users activity and compute statistics related to their physical activity.
Before exploiting \name, multiple sanitizers corresponding to various utility and privacy trade-offs are built during the training. These models are then deployed on the smartphone. During the online phase, \name selects the best sanitizer for the associated user. Both the training and the online phases are summarized in Figure~\ref{fig:archi} and explained in the following subsections.

\subsection{Building multiple sanitizers}
\label{sec:training}

The training phase is performed only once and aims at learning multiple sanitizers. 
%Seb*: il faudrait expliquer en quoi ces modèles vont différer les uns des autres
This training is performed with a reference dataset used in activity recognition, the MotionSense dataset that we describe in Section~\ref{sec:datasets}.
As shown in Figure~\ref{fig:archi}, \name is composed of multiple building blocks that we detail hereafter: 1) a sanitizer, 2) a discriminator, 3) a predictor, 4) a distortion measurement and 5) a multi-objective loss function.
%Each building bloc is explained in the following.
%With respect to our framework inspired from GANs to simplify the notation, in the rest of the paper we will refer to:
%\begin{itemize}
%\item the sanitizer as \san{}, which takes as input \dataset, the raw data, and output, $\bar{D}$, the sanitized data.
%\item the discriminator as \dis{}, which predicts the sensitive attribute $\bar{S}$ from $\bar{D}$,
%\item and will finally the \utpname{}, which predicts $\hat{Y}$ from $\bar{D}$, from will be designated by \utp{}.
%\end{itemize}   

\begin{itemize}
    \item \textbf{Discriminator:}
The discriminator $D_{isc}$ guides the sanitizer through the process of removing information related to the sensitive attribute $\sensAttr \in \{0,1\}$. 
In practice, we use a Convolutional Neural Network (CNN), which is well-suited to capture time-invariant features in time series~\cite{DLReview}.
The architecture of this CNN is presented in Appendix~\ref{sec:dnetarchi}.
The discriminator is trained to infer the sensitive information from the output of the sanitizer.
%(Section \ref{sec:sani})
The training of the discriminator is based on a loss function measuring the Balanced Error Rate \myBer{} \cite{Feldman2014} between the output of the discriminator and the ground truth sensitive attribute, which is defined as: 
%
%The \myBer{} is defined as follow:
%The $\myBer{}$ (Equation \ref{eq:ber}) is more relevant than $\mySac{}$ for datasets in which the sensitive attribute is unbalanced, as it captures the predictability of both classes.
%
%The task of the discriminator is to infer the sensitive attribute from the output of the sanitizer. 
%This neural network is implemented through 5 dense layers with both \textit{ReLU} and \textit{LeakyReLU} 
%as activation functions (the details of the architecture of the NN are presented in Annexe~\ref{sec:dnetarchi}).
%The training of the discriminator is based on a loss function measuring the Balanced Error Rate (\myBer{}). 
%This loss function aims at obtaining a \myBer{} close to $0$ which translates into a nearly perfect accuracy. 
%The \myBer{} is defined as follow:
%The $\myBer{}$ (Equation \ref{eq:ber}) is more relevant than $\mySac{}$ for datasets in which the sensitive attribute is unbalanced, as it captures the predictability of both classes.
%
%%%%%%%%%%%%%%%%%%% Thééééeoooooo: Est ce que le discriminateur prend en entrée les données corporelles ? ou bien est-ce juste les données de capteurs, qu'en est-il des données d'activités ?

{\footnotesize
%\vspace{-6mm}
\begin{align}
	BER(D_{isc}(\bar{\notSensAttr}, \decAttr), \sensAttrL{}) = \dfrac{1}{2} (\sum_{\sensAttrL{}=0}^{1}P(Disc(\bar{\notSensAttr}, \decAttr) \neq \sensAttrL{}| \sensAttr{} = \sensAttrL{})).
            \label{eq:ber}
\end{align}}

%\textbf{Définir s. where $s$ is one of the possible class values for S ?}
%This loss function aims at obtaining a \myBer{} close to $0$ which corresponds to a nearly perfect accuracy. 

The value of \myBer{} ranges between $0$ and $0.5$, in which a value close to $0$ corresponds to a perfect accuracy for the prediction of the sensitive attribute while $0.5$ means the discriminator is unable to retrieve any information about the sensitive attribute from the sanitized data. 
Hereafter, we will refer to this loss by $Loss_{Sensitive}$.\\
 
%If a classifier makes no prediction error, then the \myBer{} is $0$ and the accuracy is $100\%$. 
%On the other hand if a classifier is poor a predicting a targets, 
%the accuracy is equal to the proportion of the majority class in the dataset while the \myBer{} is $\dfrac{1}{2}$.
%The discriminator loss function is thus $Loss_{Sensitive} = \myBer{}(D_{isc}(\bar{\notSensAttr}, \bar{\decAttr}), \sensAttrL{}) $
%\begin{align}
%            Adv(\notSensAttr, \decAttr) \neq \sensAttrL{}| \sensAttr{} = \sensAttrL{})
%            \label{eq:adv}
%\end{align}
%}

\item \textbf{Predictor:}
The predictor $P_{red}$ aims at helping the sanitizer in preserving as much information as possible with respect to the activity recognition task. 
We also use a CNN for the predictor that has been optimized for predicting the user activity from the sanitized data. 
The architecture of this CNN is presented in Appendix~\ref{sec:pnetarchi}.
Thus, the predictor is trained to maximize the accuracy in inferring activities from the output of the sanitizer.
%(Section \ref{sec:sani})
We also use the balanced error rate as the loss function that should minimize the error between the output of the predictor and the ground truth of the activity: $BER(P_{red}(\bar{\notSensAttr}), \decAttrL{})$. %$BER(P_{red}(\bar{\notSensAttr}, \bar{\decAttr}), \decAttrL{})$.
For the rest of the paper, we will refer to the predictor loss as $Loss_{Activities}$.\\

\item \textbf{Distortion measurement:}
The last constraint on the sanitizer is the minimization of data distortion between the raw and sanitized data. 
Specifically, this distortion should be limited to keep as much information as possible in the sensor data for subsequent analytical tasks. 
The data distortion is measured through the $L_1$ loss function denoted \textit{l1}, applied independently on each attribute.
%L1 Loss Function is used to minimize the error which is the sum of the all the absolute differences between the true value and the predicted value.
%In other words, the distortion measurement output a vector $\overrightarrow{distortion} = \overrightarrow{\{l1(\notSensAttr_i, \bar{\notSensAttr}_i), \ldots\}}; i \in {sensor\:axis}$
%where each component is the \textit{l1-loss} applied on each sensor axis $i$. 
For two vectors $\notSensAttr_{i}$ and $\bar{\notSensAttr_{i}}$, corresponding respectively to the raw and sanitized sensor data, the loss function is defined as follows:

{\footnotesize
	\begin{align}
\vspace{-10mm}
	l1(\notSensAttr_{i}, \bar{\notSensAttr_{i}}) = \dfrac{1}{N_{\notSensAttr}} \sum_{j=1}^{N_{A}} 
	|\notSensAttrL_{ij} - \bar{\notSensAttrL}_{ij}|,
	\end{align}
}
in which $N_{A}$ is the number of possible values for a particular attribute (\textit{e.g.}, the number of axes of the accelerometer or the gyroscope), $\notSensAttrL_{ij} \in \notSensAttr_{i}$ and $i$ denotes a single observation in the window of length $T$.\\

\item \textbf{Sanitizer:}
%\label{sec:sani}
The sanitizer $S_{an}$ modifies the raw data to remove information correlated with the sensitive attribute while maintaining useful information for activity detection. Since the raw and sanitized data belong to the same space, we have implemented the sanitizer as an auto-encoder.
%To do that, the sanitizer is implemented through an auto-encoder.
In a nutshell, an auto-encoder is a neural network performing a dimension reduction of the signal to compress information before trying to reconstruct the input.
%Seb*: je suggère de mettre une référence vers un auto-encoder
The sanitizer takes into account the feedback of the discriminator, predictor and distortion measurement to output the sanitized version of the input raw data. 
More precisely, these different feedbacks are integrated into a multi-objective loss function that should be minimized. 
% In a nutshell, the sanitized is updated through a multi-objective function, while the predictor and the discriminator have their own objective (\textit{cf. Adversarial Training}).
%According to the parameters of the objective function, multiple models can be builts (one per set of parameters).
The architecture of the auto-encoder is given in Appendix~\ref{sec:snetarchi}.\\

%ab: consistency description of the nn ?

%As our framework is placed between the user and the main server predicting activities, we need to ensure that the sanitizer output
%data that still lives in the same space as the original one, ensuring that they are usable for subsequent tasks that requires such
%raw sensor information. We measure the distortion using the \textit{l1-loss} applied on a sensor-axis basis, between the original
%and the sanitized datasets. In other words, the distortion measurement output a vector $\overrightarrow{distortion} = \overrightarrow{\{l1(\notSensAttr_i, \bar{\notSensAttr}_i), l1(\bodyMeasure, \bar{\bodyMeasure}), \ldots\}}; i \in {sensor\:axis}$
%where each component is the \textit{l1-loss} applied on each sensor axis $i$. The $l1-loss$ for two vectors $\notSensAttr$ and 
%$\bar{\notSensAttr}$ is defined as $l1-loss(\notSensAttr_i, \bar{\notSensAttr}_i) = \dfrac{1}{N_{\notSensAttr_i}} \sum_{j=1}^{N_{A_i}} 
%|\notSensAttrL_{ij} - \bar{\notSensAttrL}_{ij}|$ with $\notSensAttrL_{ij} \in \notSensAttr_{i}$ and $N_{A_i}$ the number of elements 
%of $\notSensAttr_i$.

\item \textbf{Multi-objective loss function}
%As mentioned earlier, the sanitizer is implemented through an auto encoder which generates the protected data according to the a multi objectives loss function.
The multi-objective loss function $J^{S_{an}}$ drives the transformation performed by the auto-encoder to generated the sanitized data $\bar{\dataset{}}$.
This loss function takes into account three components, the capacity to detect the activity of the user (\textit{i.e.}, the output of the predictor), the capacity to detect the sensitive attribute (\textit{i.e.}, the output of the discriminator), and the level of distortion introduced in the sanitized data compared to the original one.
More formally, the multi-objective is defined as follows:

{\footnotesize
%\vspace{-4mm}
\begin{align*}
	%&\overrightarrow{J}^{S_{an}}(D, \:S_{an}, \:D_{isc}, \:P_{red)} = \overrightarrow{(\alpha * d_s(\sensAttr{}, \hat{\sensAttr{}}), \lambda * d_p(\decAttr{}, \hat{\decAttr{}}))} || \\
	%&\beta * (d_{r}(D, \:S_{an}(D))),
	% &J^{S_{an}}(D, \:S_{an}, \:D_{isc}, \:P_{red}) = \{\alpha * d_s(\sensAttr{}, D_{isc}(S_{an}(D))), \lambda * d_p(\decAttr{}, P_{red}(S_{an}(D))), \beta * (d_{r}(D, \:S_{an}(D)))\},
	J^{S_{an}}(\dataset{}, \:S_{an}, \:D_{isc}, \:P_{red}) = \{&\alpha * d_s(\sensAttr{}, D_{isc}(S_{an}(\dataset{}))), \\
	&\lambda * d_p(\decAttr{}, P_{red}(S_{an}(\dataset{}))), \\ 
	&\beta * d_{r}(\dataset{}, S_{an}(\dataset{}))\},
\end{align*}
%\vspace{-4mm}
}
in which $d_s(x)=\dfrac{1}{2} - Loss_{Sensitive}$, $d_p = Loss_{Activities}$ and $d_{r} = \{ l1(\notSensAttrL_{:,j}, \bar{\notSensAttrL}_{:,j}), \ldots\}$ with $\notSensAttrL_{:,j}$ representing a dimension of all timesteps of a single sliding window.
%Seb*: reformuler la fin de la phrase suivante qui n'est pas très claire
%, where $$d_s$$ is equal to $$\dfrac{1}{2} - BER(D_{isc}(\notSensAttr{}, \decAttr{}), \sensAttrL{})$$ on the sensitive attribute. 
The term $\dfrac{1}{2}$ in $d_s(x)$ comes from the objective of maximizing the error of the discriminator, since the sanitizer aims at sanitizing the data so that the discriminator is no more able to infer sensitive information. 

A gradient descent is applied on $J^{S_{an}}$ to minimize the global loss function following a similar approach as in \cite{aivodji2019agnostic}.
Note that each loss term is weighted with a hyperparameter. 
More precisely, $d_s$, $d_p$ and $d_r$ are weighted respectively with $\alpha$, $\lambda$ and $\beta$. 
The parameter $\alpha$ represents the relative importance given to the privacy while $\lambda$ controls the utility (\textit{i.e.}, the quality of activity detection).
As we impose the constraint that $\alpha + \lambda + \beta = 1$, we only adjust $\alpha$ and $\lambda$ hyperparameters, leaving $\beta = 1 - (\alpha + \lambda)$. 
%We believe this setup is more realistic, as end users can control the utility and privacy trade-off.

\end{itemize}

\subsection{Training Phase}
\label{sec:trainingphase}

During the training phase, we build a sanitizer for each set of possible values for the hyperparameters $\alpha$ and $\lambda$ to explore the domain of the multi-objective loss function. 
This exploration will allow \name to select the best model for each user during the online phase. 
The training procedure is summarized in Algorithm~\ref{alg:example}.

In order to optimize the utility and privacy trade-off for a specific set of $\alpha$ and $\lambda$ (line 1, Algorithm~\ref{alg:example}), the three neural networks are trained in an adversarial manner. 
This adversarial training can be seen as a game between the sanitizer on one side and the predictor and the discriminator on the other side. 
These neural networks compete against each other with opposing objectives until an equilibrium is reached. 
More precisely, the sanitizer is trained to fool the discriminator and maintained a high activity detection quantified with the predictor while limiting the data distortion. 
We follow the standard training procedure of GANs consisting in alternating in an iterative manner (at each batch of data) the training of each model with their respective loss function until convergence or until a maximum number of epoch
(\textit{i.e.}, we do not consider Competitive Gradient Descent~\cite{schafer2019competitive}).

Specifically, after initialization (lines $1-8$) the training of the sanitizer starts with $J^{S_{an}}$ 
while the discriminator and the predictor are frozen (lines $11-12$).
Once the training of the sanitizer has converged, the predictor and the discriminator are trained 
independently with their respective loss function while the sanitizer is frozen (lines $13-20$). 
These two neural networks are trained until convergence (\textit{i.e.}, until the loss no longer decreased) or if a maximum number of iterations, respectively $K_{pred}$ and $K_{disc}$, is reached. 
This two-steps process is performed iteratively until an equilibrium is reached.

\begin{algorithm}[h!]
   \caption{\name training algorithm}
   \label{alg:example}
	\begin{algorithmic}[1]
	\STATE {\bfseries Input:} $\dataset{}$, $\lambda$, $\alpha$, $max\_epoch$, $batch\_size$, $K_{pred}$, $K_{disc}$.
	\STATE {\bfseries Outputs:} $S_{an}$, $D_{isc}$, $P_{red}$.
	\STATE {\bfseries train(M, **trParams):} Train the model M using trParams.
	\STATE {\bfseries freeze(M):} Freeze the model M parameters and avoid modifications.

	\STATE \COMMENT{Initialisation}
	\STATE $S_{an}$, $D_{isc}$, $P_{red}$, $\dataset{}_d = \mathsf{shuffle}(\dataset{})$, $\dataset{}_p = \mathsf{shuffle}(\dataset{})$
	\STATE $\mathsf{Iterations}=\frac{|D|}{\mathsf{batch\_size}}$

	\STATE \COMMENT{Training Procedure}
   	\FOR{$e=1$ {\bfseries to} $max\_epoch$}
   		\FOR{$i=1$ {\bfseries to} $\mathsf{Iterations}$}
			\STATE Sample batch B of size $\mathsf{batch\_size}$ from $\dataset{}$
			\STATE train($S_{an}$, $B$, $J^{S_{an}}$, $\alpha$, $\lambda$, freeze($P_{red}$), freeze($D_{isc}$))
			
			\FOR{$k=1$ {\bfseries to} $K_{pred}$}
				\STATE Sample batch $B$ of size $\mathsf{batch\_size}$ from $\dataset{}_p$
				\STATE train($P_{red}$, $B$, $Loss_{Activities}$, freeze($S_{an}$))
			\ENDFOR
			\FOR{$k=1$ {\bfseries to} $K_{disc}$}
				\STATE Sample batch $B$ of size $\mathsf{batch\_size}$ from $\dataset{}_d$
				\STATE train($D_{isc}$, $B$, $Loss_{Sensitive}$, freeze($S_{an}$))
			\ENDFOR
	   	\ENDFOR
   	\ENDFOR
\end{algorithmic}
\end{algorithm}

\subsection{Online Phase}
\label{sec:online}

Once deployed on the smartphone, \name is composed of four components as depicted in Figure~\ref{fig:archi}: the sanitizer, the discriminator, the predictor and an activity detection component. 
Specifically, \name knows all the sanitizer, predictor and discriminator models built during the training phase. 
This set of models correspond to the different possible utility and privacy trade-offs (\textit{i.e.}, set of values explored for the $\alpha$ and $\lambda$ hyperparameters).
The selection of the model is performed by maximizing $S(P,U) = xU + yP,$
%
%
%The selection of the model is performed by maximizing the following function:
%
%$$ S(P,U) = xU + yP,$$
%
where $x$ and $y$ being positive weight coefficients with $x+y=1$, $U$ the evaluation of the activity done by the predictor, and $P$ the accuracy in terms of privacy as $ P = 1 - |0.5 - p|,$
%follows:
%
%$$ P = 1 - |0.5 - p|,$$
%
%where $p$ is the evaluation of the gender done by the discriminator, and $c$ the ideal gender accuracy corresponding to a random guess, in our case $c=0.5$.
where $P$ is the evaluation of the gender done by the discriminator. Consequently, $P$ is higher when the evaluation of the gender accuracy corresponds to a random guess (i.e., an accuracy of $0.5$). According to the expected utility and privacy trade-off, the coefficients $x$ and $y$ can be tuned (Figure~\ref{fig:selection-model}).

To find the best sanitizer over time (according to coefficients $x$ and $y$), \name evaluates the utility and the privacy of all models to select the best one. This evaluation requires to know the actual activity performed by the user and the sensitive attribute. 
While the sensitive attribute can be given by the user, the motion sensor data are not labeled with the activities as it is rather the objective of the activity recognition task to perform this inference.

We use the activity detection component (see Figure \ref{fig:archi}) to annotate some motion sensor data with their activities on the smartphone. More precisely, we ask the user to follow a specific calibration process at the installation of \name. During this process, the user is asked to perform a series of different activities for short periods to learn a specific supervised classifier to detect his activities. 
As the quantity of data available to train this classifier is limited, we rely on the use of random forests that are adapted to this context \cite{jourdan2018toward}.
This random forest (RF) classifier is then used to label the raw data in order to evaluate the utility of all sanitizers.
This evaluation is performed on a regular basis (\textit{e.g.}, each period of $p$ windows) and we compute the average accuracy over this period.
%for each data window and the accuracy of the predictor is given as the average performance of the classifier on a batch of windows.
%This evaluation is performed at each window of data and consists to measure for the last $p$ windows the average accuracy of the predictor based on the label detected by the random forest classifier, and the accuracy of the discriminator according to the sensitive attribute informed by the user.
By following this process, \name is able to identify over time the sanitizer providing the best utility and privacy trade-off defined as a measure combining the accuracy of the activity recognition and the inference of the sensitive attribute.

\section{Experimental setting}
\label{sec:experiment} 

%The main objective of our experiments is to evaluate the capacity of \name to sanitize sensor motion data to prevent sensitive attribute inference while limiting the loss of information with respect to activity recognition.
%In this section, we present the datasets used to evaluate the performance of \name (Section~\ref{sec:datasets}), %the methodology used to train the neural network models (Section~\ref{sec:methodo}), 
%the baselines we compared with (Section~\ref{sec:baselines}), the evaluation metrics considered (Section~\ref{sec:metrics}) as well as the methodology adopted (Section~\ref{sec:methodo}).

\subsection{Datasets}
\label{sec:datasets}

We used two real datasets, which are both publicly available and heavily used in the literature: MotionSense and MobiAct.

\begin{itemize}
    \item \textbf{MotionSense}~\cite{ortiz2015smartphone} contains data captured from an accelerometer (\textit{i.e.}, acceleration and gravity) %, 
%altitude (i.e., pitch, roll and yaw) 
and gyroscope at a constant frequency of 50Hz collected  with an iPhone 6s kept in the participant's front pocket.  
Overall, a total of 24 participants have performed six activities during 15 trials in the same environment and conditions.
%Seb: je suggère de rajouter la longueur de chaque essai
The considered activities are going downstairs, going upstairs, walking, jogging, sitting and standing. 
%In addition, attributes of the participants such as gender, age, weight and height have also been collected.
%The specificity of this database with respect to others is that the signals are relatively noisy as they were acquired through the smartphone kept in a pocket.

\item{\textbf{MobiAct}}~\cite{vavoulas2016mobiact} records the data from 58 subjects during more than 2500 trials, all captured with a smartphone in a pocket.
This dataset includes signals recorded from the accelerometer and gyroscope sensors of a Samsung Galaxy S3 smartphone with subjects performing nine different types of activities of daily living. 
For our experiments, we only used the trials corresponding to the same activities as the MotionSense dataset.
\end{itemize}
%All the datasets are split in trials with 2/3 of trials for training/validation and 1/3 for testing.

%Both datasets are publicly available and heavily used in the literature. These 
Both datasets are balanced and contains an equal number of males and females. 
%In addition, all activities are represented with the same quantity of data. 
%Seb: attention il faudrait répondre à la phrase ci-dessous
%\textbf{Il me semblait qu'il y avait plus de marche ?}
%Seb: ci-dessous on pourrait préciser si la séparation training et testing se fait de manière aléatoire ou selon un critère particulier par exemple au niveau chronologique
The datasets are split between training and testing, with 2/3 of trials used for training and validation and 1/3 for testing.
These two datasets share similar characteristics, which allows to test the transferability of the models from one dataset to the other. More precisely, the models learned on one dataset can be used to sanitize data from the other dataset.
This evaluation corresponds to a more realistic use case and to the best of our knowledge was never considered in previous work related to the sanitization of sensor data.

%\subsection{Methodology}
%\label{sec:methodo}

%Do we have something to put here?

%\subsubsection{How to choose $\alpha$ and $\lambda$ ?}
%We have noticed that usually very few optimisation methods are done to tune the hyperparameters of the model. Here we based our optimisation on the building of a L-curve framework \cite{Belge_2002}.

%\subsubsection{Early stopping and dynamic learning rate}
%In the model, two parameters are integrated to control the number of iterations that the Predictor ($K_pred$) and the Discriminator ($K_disc$) do for one iteration of the Sanitization. It reflects how much the Predictor and Discriminator are re-trained for each new sanitized data produced.

%We decided to not use a constant number of iterations for the training of the model, early stopping is a method that consider the test loss in order to stop the training session when a given threshold is reached. Moreover, this threshold have to be reached during a certain number of epoch to stop the training. In our case, we use the parameter :

\subsection{Baselines}
\label{sec:baselines}

To assess the performance of \name, we considered a set of baselines that we detail hereafter. 
One of these baselines is based on a random forest classifier \cite{jourdan2018toward} while the others are based on GANs \cite{Malekzadeh2018,olympus_raval,Malekzadeh_2019}. 
Regarding GAN approaches, authors use an architecture of neural networks slightly different to ours.
%Seb: ci-dessus il faudrait dire qui sont "they"
To provide a fair comparison, we propose to implement their functionalities in our architecture (number of layers, type of CNN, \ldots).
%Seb*: la phrase ci-dessus ainsi que celle ci-dessous ne sont pas très claires et devrait être reformulée
This methodology allows to assess the main characteristics adopted in the baselines without depending on their choice of architecture that can also have an impact on performance.
%\textbf{parler de l'évolution des baselines ... }

\begin{itemize}
    \item \textbf{ORF:}
%\subsubsection{ORF}
To limit the exposure of the data, in~\cite{jourdan2018toward} the raw data is preprocessed on the user's smartphone and only relevant features are transmitted to the application hosted on the cloud. 
The relevant features are first identified according to the target application (\textit{e.g.}, activity recognition) and selected either in the temporal or the frequency domain.
Originally proposed to avoid users re-identification, we adapt this approach to prevent the inference of the sensitive attribute, namely gender. 
More specifically, we first detect the features that are the most correlated with the gender before normalizing the features in the frequency domain and removing the features in the temporal domain that are not used for the activity classification.\\

%\subsubsection{GEN}
\item \textbf{GEN:}
Similarly to \name, GEN (Guardian Estimator Neutralizer)~\cite{Malekzadeh2018} also relies on an adversarial approach to optimize the utility and privacy trade-off.
However, this solution does not follow the standard iterative training procedure of GANs as described in Section~\ref{sec:trainingphase}.
More precisely, the first network, a classifier, is learned once on the raw data to identify both sensitive (\textit{e.g.}, the gender) and non-sensitive information (\textit{e.g.}, the activity).
Then the second network, an auto-encoder, is also trained only once through a loss function that does not take into account the data distortion.
Finally, the model used in the online phase is the same for all users and corresponds to the best set of hyperparameters identified during the training phase. 
%Seb: j'ai du mal à comprendre pourquoi on utilise pas leur architecture pour se comparer, en effet un reviewer pourrait dire que la comparaison ne fait pas de sens car on a pas utilisé la méthode originelle proposée par les auteurs
While this solution relies on a neural network architecture slightly different from ours, we implement GEN by using our architecture. 
However, to evaluate the performance of GEN in a context of transfer learning, we also use their original neural networks (learned on MotionSense\footnote{\url{https://github.com/mmalekzadeh/motion-sense/tree/master/codes/gen_paper_codes}}) to assess its performance on MobiAct.\\
%\textbf{Ne faut-il pas aussi parler de leur implémentation qu'on utilise également ?}

\item \textbf{Olympus:}
%\subsubsection{Olympus}
This approach~\cite{olympus_raval} is similar to GEN 
%\textbf{Pourquoi se comparer à GEN ? Il faut se comparer à nous !}
with the exception that two different neural networks are used to learn the sensitive attributes and to learn non sensitive information. 
%\textbf{2 fois sensitive dans la phrase précédente. Est-ce normal ?}
In addition, these classifiers are trained using sanitized data by following an iterative process similar to \name described in Section~\ref{sec:trainingphase}.
However, the loss function does not account for data distortion and the model deployed is the same for all users.
While this approach is used for a different objective (\textit{i.e.}, to avoid users re-identification), 
we adapt it by using our architecture.\\

%\subsubsection{MSDA}
\item \textbf{MSDA:}
This solution~\cite{Malekzadeh_2019} can be viewed as an evolution of Olympus 
%\textbf{Pourquoi se comparer à Olympus ? Il faut se comparer à nous !} 
in which the loss function driving the training of the auto-encoder accounts for data distortion.
However, the model used in the online phase is still the same for all users.
While this approach was originally developed with a different purpose in mind (\textit{i.e.}, to avoid re-identification), 
we adapt this solution by using our architecture.
This baseline is the closest to \name but without the dynamic sanitizing model selection in the online phase.
%This baseline represents  \name without dynamic sanitizing model selection in the online phase.

\end{itemize}
\subsection{Evaluation metrics}
\label{sec:metrics}

We evaluated \name along both utility and privacy metrics, and a couple of system-level metrics.  %(to be fixed according to metrics introduced in section 3)

\begin{itemize}
\item \textbf{Utility:}
%The utility of the sanitized data can be of different nature and is application dependent.
In our context of physical activity monitoring, the first considered utility metric is the accuracy of a classifier for activity recognition.
% To do that, we measure the accuracy of a classifier trained to detect the users activities.
More precisely, we use the confusion matrix derived by this classifier to measure the number of correct predictions made by the classifier over all predictions made. 
The value of the accuracy ranges from $0$ to $1$, in which $1$ corresponds to perfect accuracy.
%Seb: je suggère de mentionner quel serait l'accuracy si on avait un baseline qui prédit la classe majoritaire
%
In addition, analytics applications monitoring physical activity usually compute and present many estimators to users.
To evaluate this aspect, we compute the number of steps detected from the sanitized data and compare it with the number of steps in the raw data.
To realize this, we first normalize the raw and sanitized data %in order 
to compare them in the same range of values, %before fixing a threshold for each window of the raw data corresponding to the mean plus the standard deviation of the signal in the window. 
%We count the number of peaks above the threshold in both the raw data and its corresponding sanitized version.
and then compute a Peak Acceleration Threshold~\cite{7962470} from the raw data 
to estimate the number of peaks. 
More precisely, we used Adaptiv: An Adaptive Jerk Pace Buffer Step Detection Algorithm (\url{https://github.com/danielmurray/adaptiv}) for estimating the number of steps detected by the analytics application from the received data.\\

\item \textbf{Privacy:}
To assess the level of privacy of \name, we rely on the accuracy of inferring the sensitive attribute (\textit{i.e.}, the gender).
% Déjà parlé d'accuracy plus haut.
% The accuracy is comprised within $0$ and $1$.
In our case, an accuracy of $0.5$ corresponds to a random guess as our dataset is balanced. \\

%\item \textbf{Performance:}
\item \textbf{System-level:}
To assess the overhead of operating \name on a smartphone, we measure both the CPU time spent to sanitize the raw data on the smartphone and the energy consumption over time during a real-time processing of \name.

\end{itemize}

\subsection{Methodology}
\label{sec:methodo}

%Datasets are split in trials with 2/3 of trials for training and validation, and 1/3 for testing.
\name is trained only with the MotionSense dataset while the results reported for MobiAct evaluate the transfer learning (\textit{i.e.}, using sanitizing models trained on MotionSense to sanitize data from MobiAct).
In the training phase, we explore a range of values between 0.1 and 0.9 with a 0.1 step for both $\alpha$ and $\lambda$, which corresponds to 36 different sanitizing models. 
The sanitizer models of \name are trained for 300 epochs and the size of a data batch is set to 256 samples.
In the online phase, we select a privacy and utility trade-off focusing primarily on privacy (\textit{i.e.}, ensuring the protection of the gender at the cost of the accuracy). 
%Seb: ci-dessous, je suggère de nommer les paramètres
This trade-off is controlled by the parameters $x$ (utility) and $y$ (privacy) (Section~\ref{sec:online}) which are set respectively to 0.1 and 0.9.

The random forest classifier applied during the online phase of \name uses a feature vector extracted from the raw signal. The choice of these descriptors was made on the basis of an earlier review on effective descriptors for gait recognition~\cite{sprager2015inertial}.
We use 4-fold cross-validation in which the testing set is randomly partitioned into 4 equal sized subsamples.

Reported results correspond to average over 10 repetitions of each experiment.
The computation of the different global models (each corresponding to a precise set of hyperparameters) has been parallelized on a hybrid GPU/CPU computing farm.
%\textbf{should we need to add something here? on the evaluation over time?}

%To assess the level of privacy of \name, we rely on the accuracy of the sensitive attribute as well as the \myBer{} (cf. section \ref{sec:overview}),
% both computed a set of classifiers trained on the sanitized signal. As previously mentioned, the \myBer{} ranges from $0$ to $0.5$ and the accuracy is comprised within $0$ and $1$.
% On one hand, a \myBer{} value of $0$ means the classifier has not made any prediction mistake (a perfect accuracy of $1$),
% on the other hand, a value of $0.5$ correspond to the inability of the classifier to accurately predict the specified target.
% It translate into an accuracy of $0.5$ in case of balanced dataset, and into the proportion of the most represented target value otherwise.

% The objective is therefore to increase the \myBer{} value of each classifier, making it as close as possible to $0.5$, while we decrease their accuracy to $0.5$ since the dataset is balanced.

\section{Evaluation}
\label{sec:results} 

In this section, we report the results obtained for the evaluation of \name by highlighting important features, namely the good utility and privacy trade-off (Section~\ref{sec:resultsInference}), the low distortion of the sanitized data (Section~\ref{sec:resultsDistortion}), the better performances compared to state-of-the-art approaches (Section~\ref{sec:resultsComparison}), the advantage of dynamically select the best sanitizing model according to the incoming data (Section~\ref{sec:resultsDynamic}), and the limited cost of operating \name on a mobile (Section~\ref{sec:cost}). %, and the limited risk of information leakage in model selection (Section~\ref{subsec:leakage}). %and the capacity of the online phase to select the best sanitizing model for each user (Section~\ref{sec:consistency}).

%, and the comparison with baselines (Section~\ref{sec:resultsComparision}). % and the transferability to new users (Section~\ref{sec:resultsTrans}). 
%Note that MotionSense dataset has been used in experiments related to Sections~\ref{sec:resultsInference}, \ref{sec:resultsDynamic}, \ref{sec:resultsComparision} and MobiAct dataset in Section~\ref{sec:resultsTrans}.
%Our implementation is publicly available~\footnote{\name:\url{https://github.com/DynamicSanitizer/DySan}}. 
%
%Les codes doivent être rendus dipos. Je propose de ne pas le mettre.
% c'est obligatoire pour cette conf et une pratique hautement encouragée pour la plupart des autres.

\subsection{Utility and privacy trade-off}
\label{sec:resultsInference}

In this section, we evaluate the capacity of an analytics application to infer the gender of the user and its activity from the sanitized data provided by \name and sent by the mobile application.
We compare the performance of several classifiers that could be used by the analytics application, namely a gradient boosting classifier (GB), a multi-layer perceptron (MLP), a long short-term memory neural network (LSTM), a decision tree (DT),  a random forest (RF), a logistic regression (LR) and also two CNNs with the same architectures than the predictor and the discriminator of \name.

\begin{figure}[t]
	\centering
	\subfloat[MotionSense]{
	\includegraphics[width=6cm]{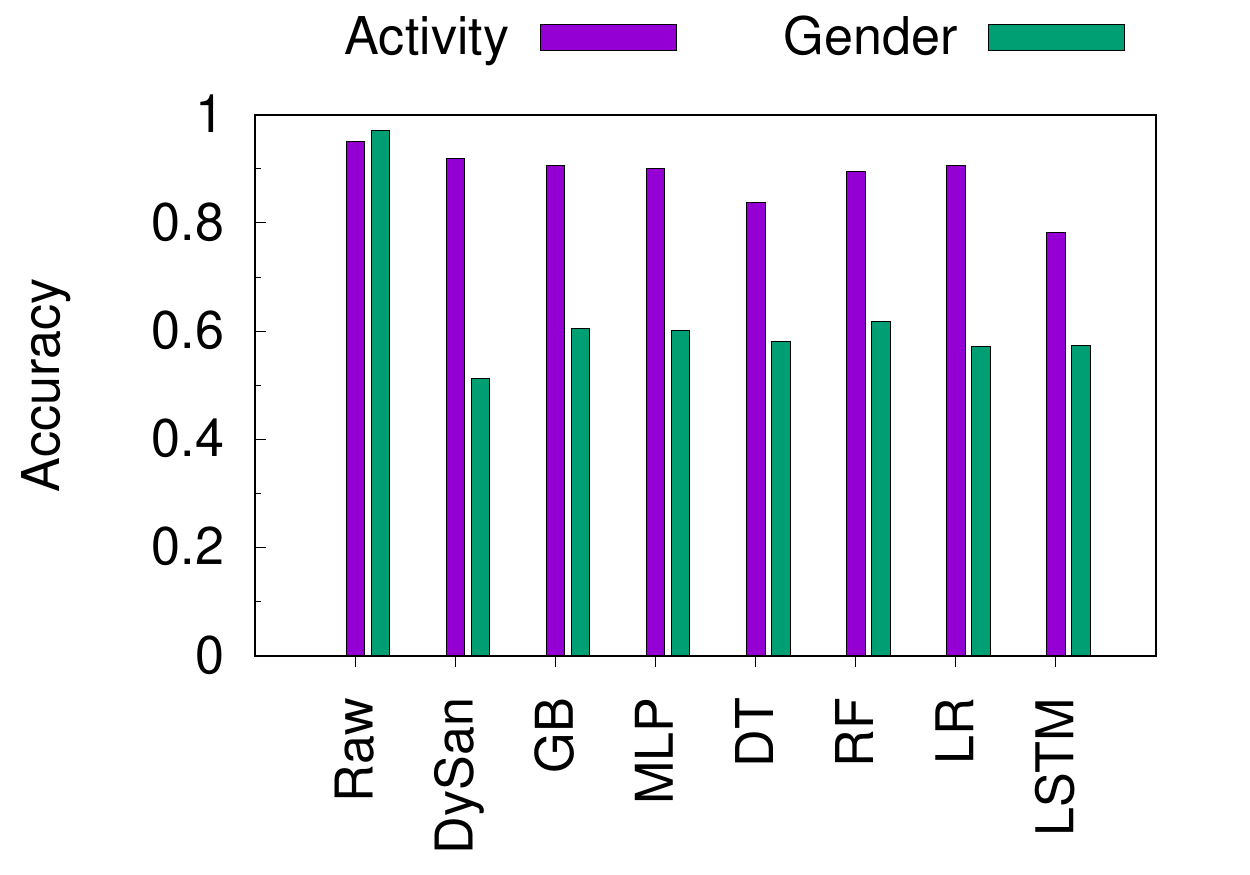}
	\label{fig:inference_ms}
	}
	\subfloat[MobiAct]{
	\includegraphics[width=6cm]{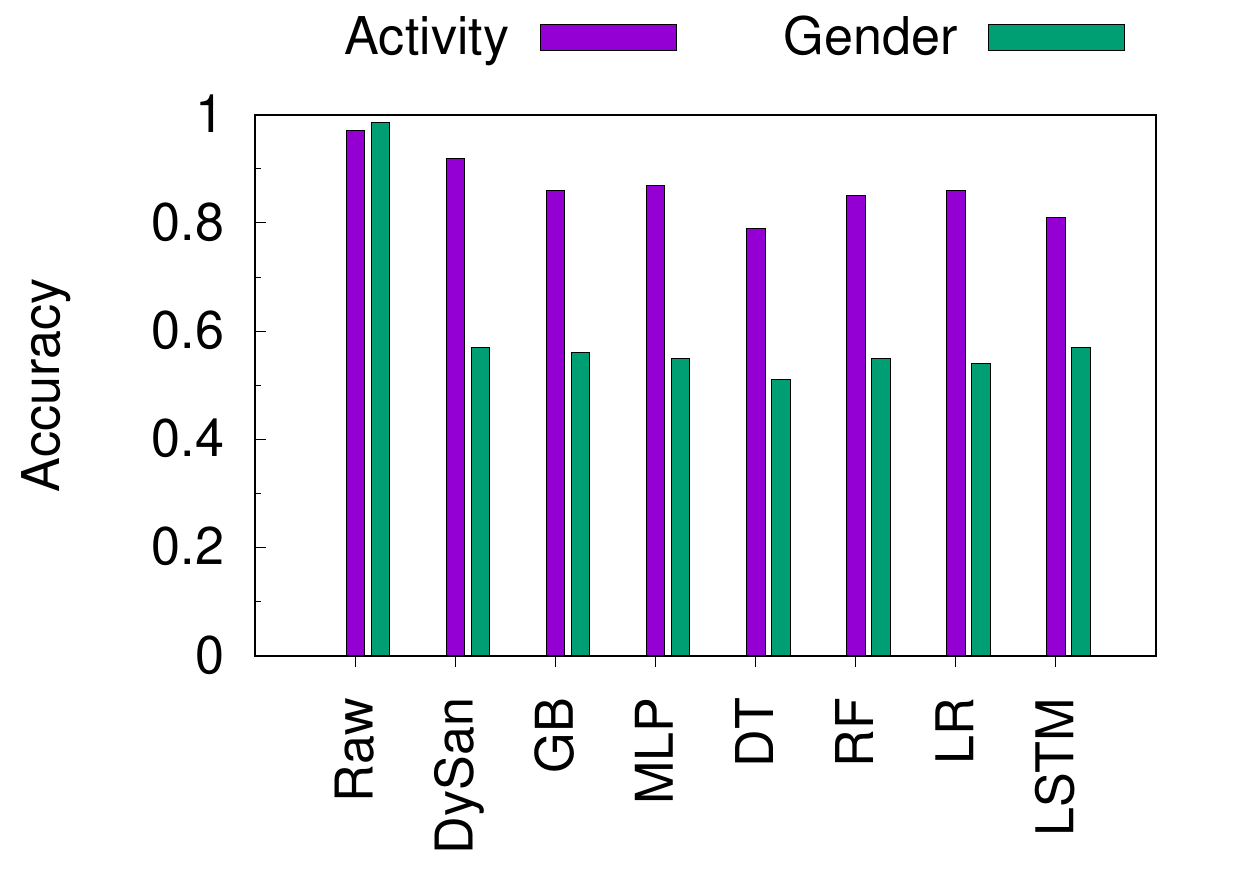}
	\label{fig:inference_ma}
	}
	\vspace{-3mm}
	\caption{\small The sanitized data provided by \name drastically decreases the privacy risk compared to using the raw data while limiting the loss of activity detection, and this regardless of the classifier used.}
	\label{fig:inference}
\vspace{-4mm}
\end{figure}

Figure~\ref{fig:inference} reports the accuracy for both datasets for predicting the gender and the activity with the different classifiers as well when using the raw data.
First, the results show that without any protection (\textit{i.e.}, on raw data) the application is able to infer the gender with 98.5$\%$ accuracy, %98$\%$ of accuracy, which represents an important privacy risk. 
In addition, the activity is also inferred from the raw data with 97$\%$ of accuracy on average. %95$\%$ 
Secondly, we can observe that \name successfully decreases the privacy risk with respect to inferring the sensitive attribute while limiting the loss of activity detection.
Indeed, with the sanitized data, an analytics application is only able to infer the gender up to 61$\%$ and 57$\%$ of accuracy, respectively for MotionSense and MobiAct.
In term of utility, depending on the classifier, the accuracy of the activity recognition varies between 78\% and 92\%, which represents only a small drop compared to using the use of the raw data. 
Remark that the LSTM, a recurrent neural network architecture commonly used for temporal signal, does not provide best results as one could expect.
%In term of utility, depending on the classifier, the accuracy of the activity recognition varies between 90\% and 86\% which represents only a small drop compared to using raw data.

%We evaluate here the contribution of \name by comparing the utility and privacy between raw data and sanitized data. To do so, we compare the performance of several classifiers that could be hosted in the cloud and be learned through an analytics application. In our case, we tested a gradient boosting classifier (GB), a multi-layer perceptron (MLP), a decision tree (DT), a random forest (RF), a logistic regression (LR) and also two CNNs with the same architectures than the Predictor and the Discriminator (DySan). 
%Figure~\ref{fig:inference} shows that without protection (on raw data) we have 98$\%$ in gender classification, which means high privacy risks. When any classifier presented before is trained and tested on sanitized data we decrease this privacy risk at 60$\%$. In the same time, the activity recognition is maintained above 85$\%$ of accuracy for DT and reaches 92$\%$ for DySan.

%\textbf{plot to be updated with last results}

\subsection{Distortion of the sanitized signal}
\label{sec:resultsDistortion}

The utility of the sanitized data is not just about the activity recognition but also with respect to more fine-grained information related to the activity. 
In this section, we demonstrate that \name keeps relevant information in the signal enabling to conduct further analysis.
More precisely, we consider the computation of the number of steps from the signal for MotionSense dataset. 
Following the step detection method presented in \ref{sec:metrics}, Table~\ref{tab:step} shows that with \name the estimation of the number of steps only suffers from a 7$\%$ error compared to the raw data. 
With the different baselines, the sanitized signal appears to be much more noisy and the step detection is greatly impacted with an overestimate number of the steps of more than 64$\%$ for Olympus, more than 29 $\%$ for MSDA and more than 12$\%$ of errors for GEN.
The method ORF is not considered here because it only extracts features and the signal is not preserved, which prohibits possibility to conduct further analysis.

\begin{table}[t]
\center
%\footnotesize
\begin{tabular}{|l|c|c|}
  \hline
  & Steps & DTW\\
  \hline
   Raw data & 14387  & - \\
   \hline
   \name & 15321 \textbf{(+6.49 $\%$)} &   12.96 \\   
    \hline
    GEN & 12817 \textbf{(-12.25$\%$)} &  14.28  \\
    \hline
    Olympus & 23658 \textbf{(+64.44$\%$)} &  156.03  \\
    \hline
    MSDA & 18624 \textbf{(+29.45$\%$)} &  23.37  \\
  \hline
\end{tabular}
\vspace{-1mm}
\caption{\small The sanitized signal provided by \name appears to be less distorted and more useful for step detection than other approaches.}
\vspace{-0.3cm}
\label{tab:step}
\end{table}

%Seb: suggestion: mentionner cet métrique dans la section 4
To evaluate the deformation of the signal, we also report the Dynamic Time Warping (DTW)~\cite{dtw} between the raw and the sanitized data from each baseline (Table~\ref{tab:step}).
This metric measures the distortion between two temporal signals.
If this metric has a small value then it means that the two signals are quite similar to each other, which is a sign of a small distortion.
The results obtained show that the sanitized data produced by \name is more similar to the raw data compared to other baselines.
Similarly to step detection, the sanitization process of Olympus depicts a large data distortion making further analysis of the signal impossible.
Other metrics assessing the deformation of the signal (\textit{i.e.}, mean, standard deviation, skewness, kurtosis, and energy) are reported in Appendix~\ref{sec:deformation}.

%The utility of the sanitized data is not just about the classification accuracy in activities but also about more precise information on the activity level. Then to show that \name allows to keep relevant information in the signal allowing to conduct further analysis, we consider here the computation of the number of steps from the signal. Following the step detection method presented in \ref{sec:metrics}, Table~\ref{tab:step} shows that with \name we can find almost all the steps detected from raw data with less than 5$\%$ of errors. We compared this result with the two baselines implemented Olympus and MSDA (see \ref{sec:training}): the sanitized signals appears to be much more noisy and the step detection is greatly impacted with more than 64$\%$ steps overdetected for Olympus and more than 29 $\%$ for MSDA. We couldn't count steps with ORF method because it implies features transformation and the signal is not preserved.

%Ajouter un paragraphe sur la conservation des gaps entre les info permettant d'identifier les activités.

%Ici compléter les chiffres, sur MotionSense ou MobiAct ?

\subsection{Comparative analysis}
\label{sec:resultsComparison}

\begin{figure}[t]
	\centering
	\subfloat[MotionSense]{
	\includegraphics[width=6cm]{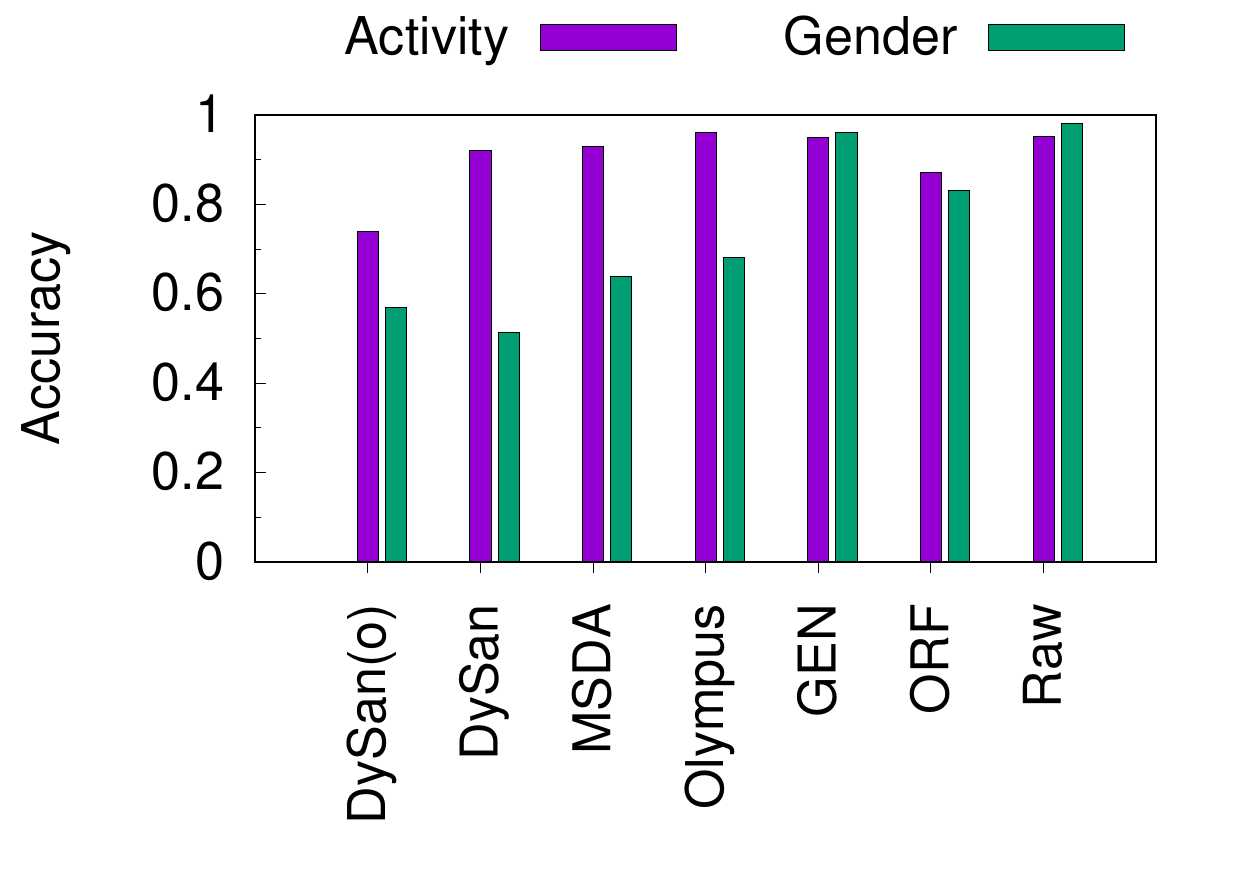}
	\label{fig:comparison_ms}
	}
	\subfloat[MobiAct]{
	\includegraphics[width=6cm]{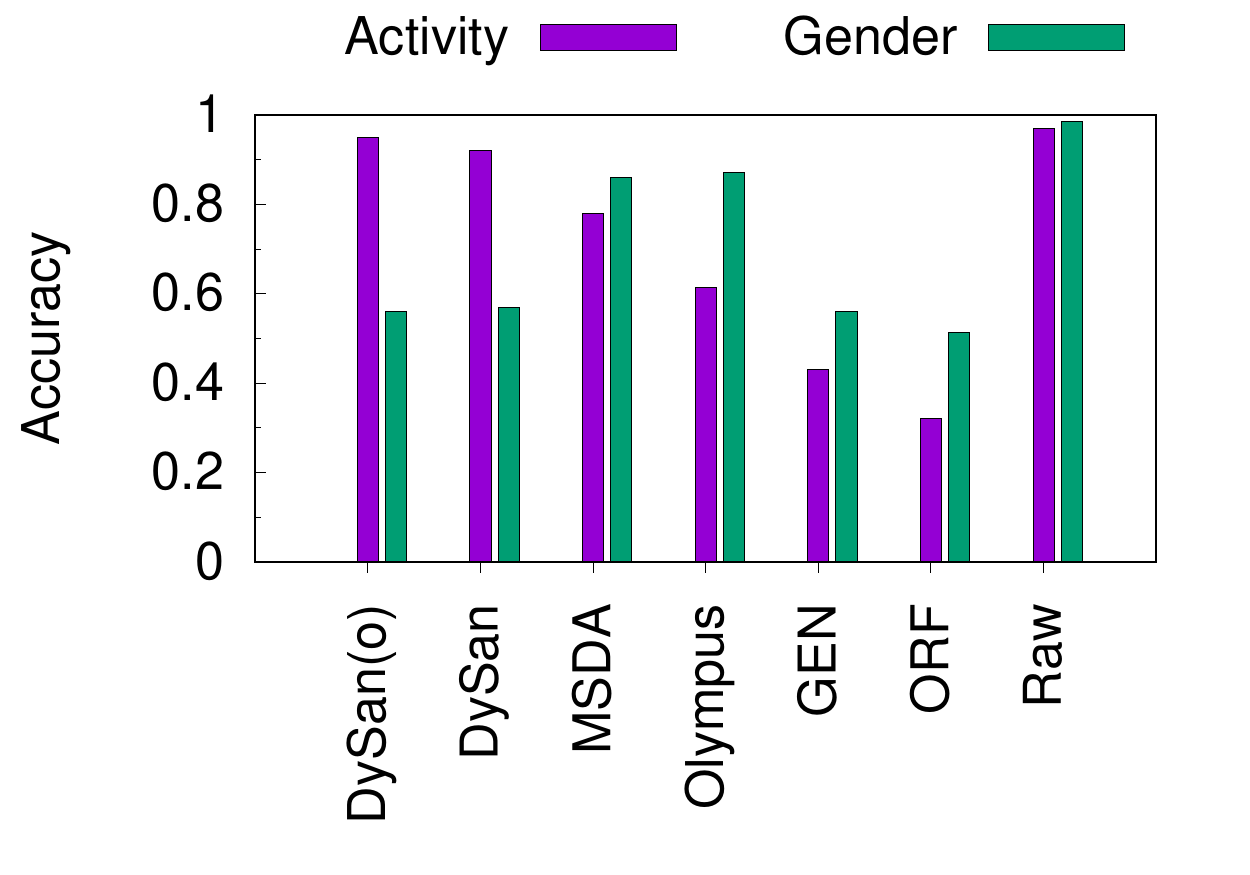}
	\label{fig:comparison_ma}
	}
	\vspace{-3mm}
	\caption{\small \name provides the best privacy protection compared to state-of-the-art approaches at the cost of a slightly 
	smaller accuracy in term of activity detection.}
	\label{fig:baseline}
\vspace{-4mm}
\end{figure}

%In this section, w
We compare \name against baseline approaches (Figure~\ref{fig:baseline}). Two versions of \name are given to represent, \name where the annotations of the activities are known and the online version, \textsc{DySan}(o), where the activities are not given but inferred from the random forest (RF) classifier. The first version has been added for a more fair comparison to state-of-the-art that does not evaluate models as we suggest.

For MotionSense (Figure~\ref{fig:comparison_ms}), the privacy improvement of \name occurs at the cost of a slightly decrease of utility (gender inference limited to 51\% and an activity recognition of 92\%). For the online version, which works blindly without annotations, the performance is a little worse, with a gender inference of 57\% and accuracy in activity of 75\%.
This utility mitigation comes from the imperfect accuracy of the random forest classifier used in the online phase to select the best sanitized model.
Indeed, to dynamically select the sanitizer model, \name needs to estimate the model providing the best utility and privacy trade-off with respect to the considered parameters (Section~\ref{sec:online}). 
To achieve this, \name relies on a calibration process to build a RF classifier on the raw data used as a reference to predict the current activity performed by the user. This RF classifier provides an average accuracy of respectively 96$\%$ and 94$\%$ on the activity recognition for MotionSense and MobiAct datasets.
While these accuracies are high, an activity wrongly predicted by this classifier leads to a selection of the sanitizer model that does not correspond to the best utility and privacy trade-off.
%We report this impact in Figure~\ref{fig:baseline}, in which the white part of the histogram reports the fall of accuracy in activity during the online phase for \name. 
%By using a perfect classifier (\textit{i.e.}, the actual annotated activity), the activity recognition could be inferred with 92$\%$ of accuracy for MotionSense and 95$\%$ for MobiAct (versus respectively 74$\%$ and 92$\%$ using our classifier). 
%The gender inference is also impacted by this imperfect prediction with a slight increase of the accuracy (from 51$\%$ to 57$\%$ for MotionSense and from 56$\%$ to 57$\%$ for MobiAct). %It should be noted that this evaluation framework is much stricter than for the other models where the annotations of the activities are systematically given. Ce qui explique les très bonnes performances en MobiAct.

As depicted on Figure~\ref{fig:comparison_ma}, results for MobiAct show that \name and \textsc{DySan}(o) outperform other approaches by limiting the gender inference to 55\% and 54\% while only reducing the accuracy of activity recognition by 2\% and 5\% compared to using the raw data, respectively. 
Although GEN and ORF also limit significantly the gender inference, the accuracy of the activity detection is drastically impacted (43\% and 32\%, respectively).

We detail in Appendix~\ref{sec:classificationDetails} the accuracy for each activity. 
From these results, we can observe that the less represented activities are the least well recognized (\textit{i.e.}, the dataset is unbalanced with more data related to the walk). 
%As MobiAct dataset contains much more data than MotionSense, the associated training captures more information on signals, leading to a better accuracy.

Results also show the performance improvement provided by each baselines approach based on adversarial networks. Specifically, GEN, Olympus and MSDA gradually improve the utility and privacy trade-off. However, our utility analysis (Table~\ref{tab:step}) shows that the sanitized data is very distorted, which harms the possibility to perform signal processing for further analysis. MSDA integrates the data distortion in its loss function, which leads to less distorted data. This feature improves the quality of signal processing but does not significantly improve the trade-off between utility and privacy compared to Olympus (Figure~\ref{fig:baseline}). 
By dynamically selecting the best sanitizer model for each window of raw data, \textsc{DySan}(o) makes the gender inference close to a random guess while preserving an accurate activity detection.

The results of GEN reported in~\cite{Malekzadeh2018} mention an accuracy of 94\% for the activity recognition and 64\% for the gender inference for MotionSense dataset compared to 95\% and 96\%, respectively in our experiments.
This difference comes from our implementation that does not used exactly the same neural network setting as the original baseline (only one neural network for both classification tasks versus two neural networks as explained in Section~\ref{sec:baselines}). 
%Seb: la phrase ci-dessous n'est pas très claire, je suggère de la reformuler
However, this difference also tends to assume an over adaptation of the underlying neural network to the considered dataset.
This over adaptation is also pointed by the complete different trend for the accuracy provided for MobiAct compared to MotionSense.
%our evaluation of transfer learning Section~\ref{sec:resultsDynamic}).

As described in Section~\ref{sec:online}, the best sanitizer model is selected according to the definition of the utility and privacy trade-off defined by weight coefficients $x$ and $y$ in the online phase. 
The reported results correspond to a privacy and utility trade-off controlled by parameters $x=0.1$ and $y = 0.9$ (Section~\ref{sec:methodo}). 
Appendix~\ref{sec:tradeoff} depicts the evolution of this trade-off according to these parameters.
%to the choice of these parameters.

% that approaches 
%based on adversarial networks provides the best utility and privacy trade-off.
%Moreover, we observe that \name outperforms other approaches 
%by improving the privacy while maintaining a high utility.
%For instance compared to MSDA, \name reduces by 12$\%$ the accuracy in 
%inferring the gender while disturbing only by 2\% the accuracy in detecting the activity.

%GEN 0.94 0.64 vs GEN : activity = 0.95, gender = 0.96

\subsection{Dynamic selection of sanitizing model} %Est-ce bien le terme dynamic ?
\label{sec:resultsDynamic}

During the training phase, \name computes the sanitizer models corresponding to all
possible utility and privacy trade-off by exploring the range of values for the hyperparameters $\alpha$ and $\lambda$. 
We evaluate here the benefit to dynamically adapt the sanitizing model according to the incoming data of each user compared to two static baseline approaches.
Firstly, we compute the accuracy for both the gender inference and the activity recognition when the sanitizer model is fixed for all the users.
This case represents the behaviors of all comparative baselines where the considered model is the one providing the best performance (\textit{i.e.}, the utility and privacy trade-off) on average for all the users. Secondly, we consider a personalized solution where the sanitizer model is personalized for each user.
In this case, the sanitizing model is the one which provides the smallest accuracy in term of gender inference and the best accuracy in term of activity recognition according to the whole models set for a specific user.
This solution provides a sanitizer model personalization but the selected model is static and does not change according to the evolution of the incoming data (and the associated changes in term of performed activity).

\begin{figure}[t]
	\centering
	\subfloat[MotionSense]{
		\includegraphics[width=0.5\linewidth]{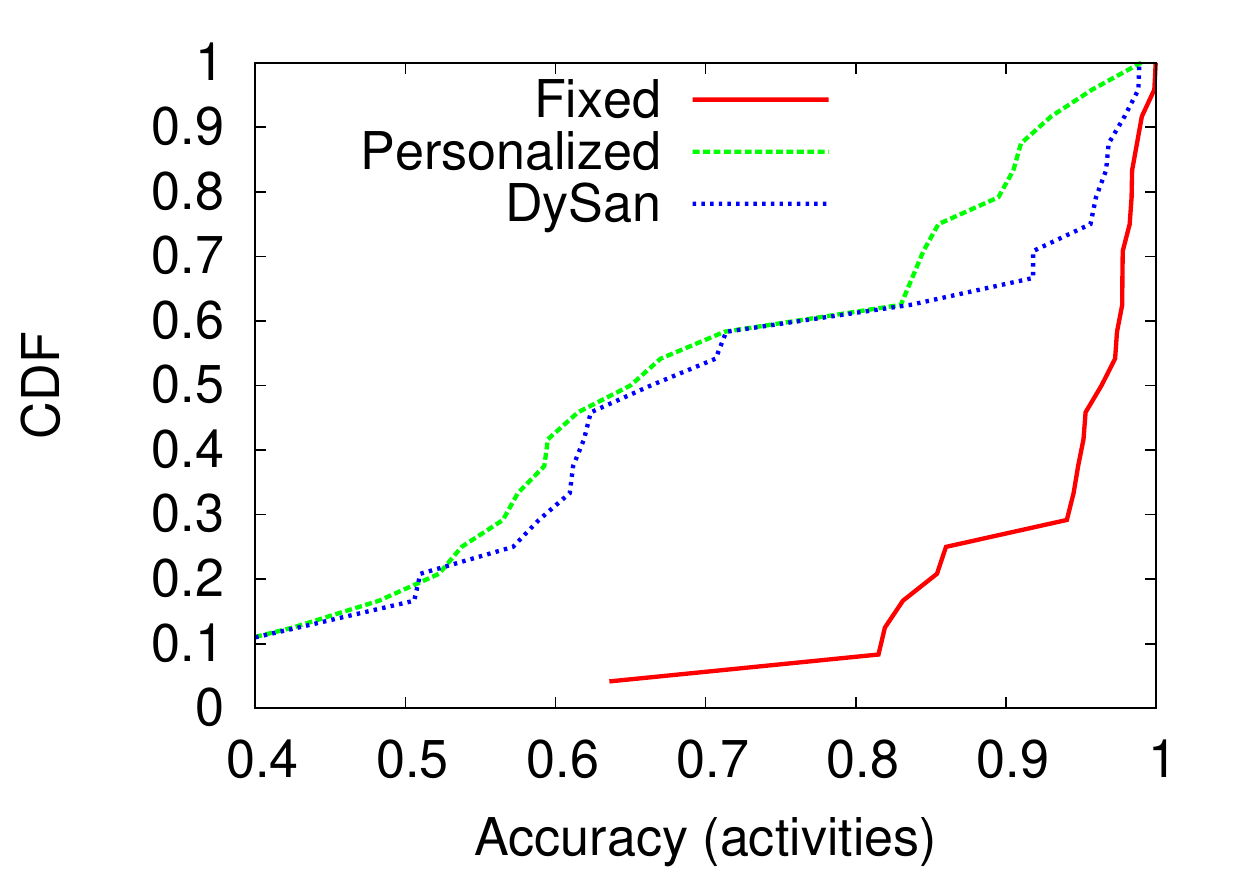}
		\label{fig:inference1MS}
	} 
	\subfloat[MobiAct]{
		\includegraphics[width=0.5\linewidth]{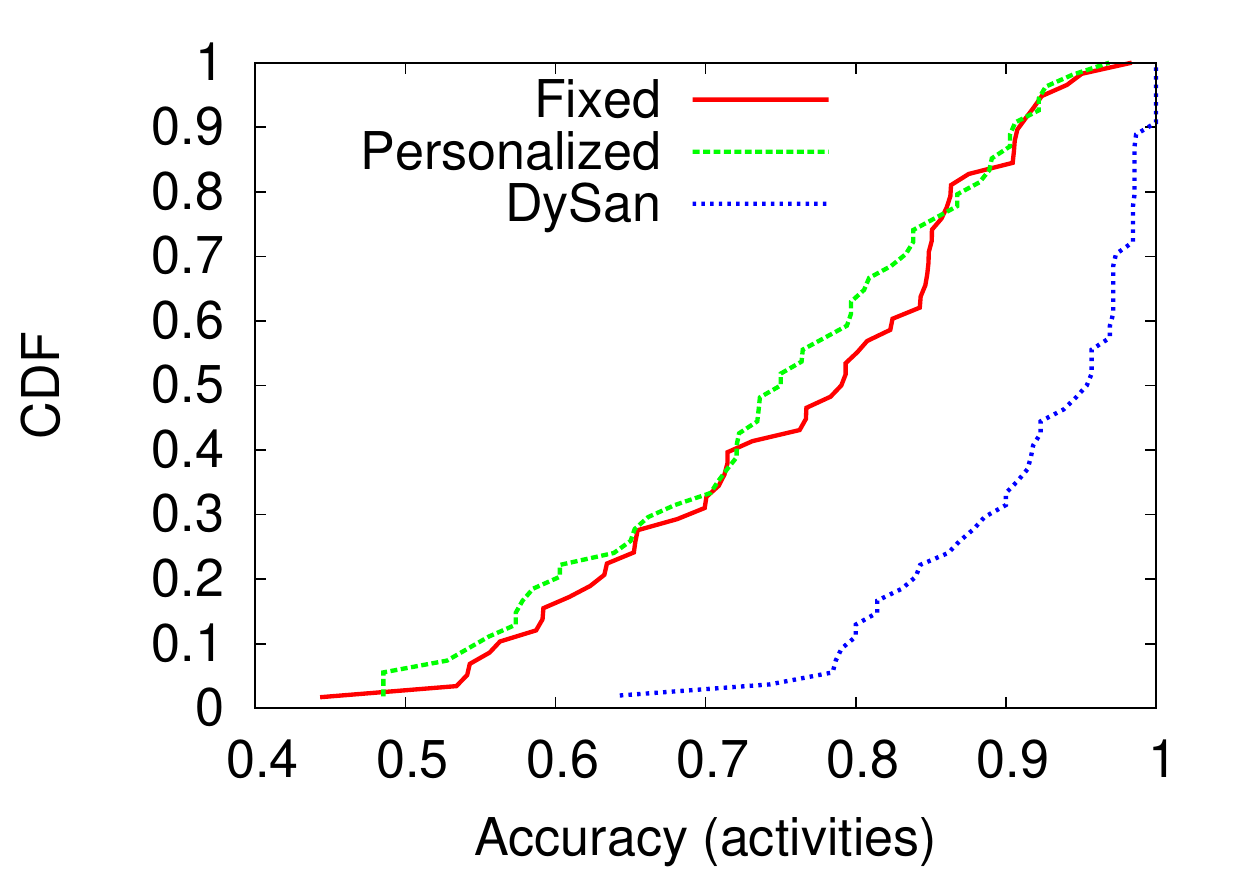}
		\label{fig:inference1MA}
	}
	\vspace{-3mm}
\caption{\small The dynamic sanitizing model selection of \name significantly improves the activity recognition in case of transfer learning (\textit{i.e.}, MobiAct dataset).}
\label{fig:cdf-activity}
\vspace{-4mm}
\end{figure}

\begin{figure}[t]
	\centering
	\subfloat[MotionSense]{
		\includegraphics[width=0.5\linewidth]{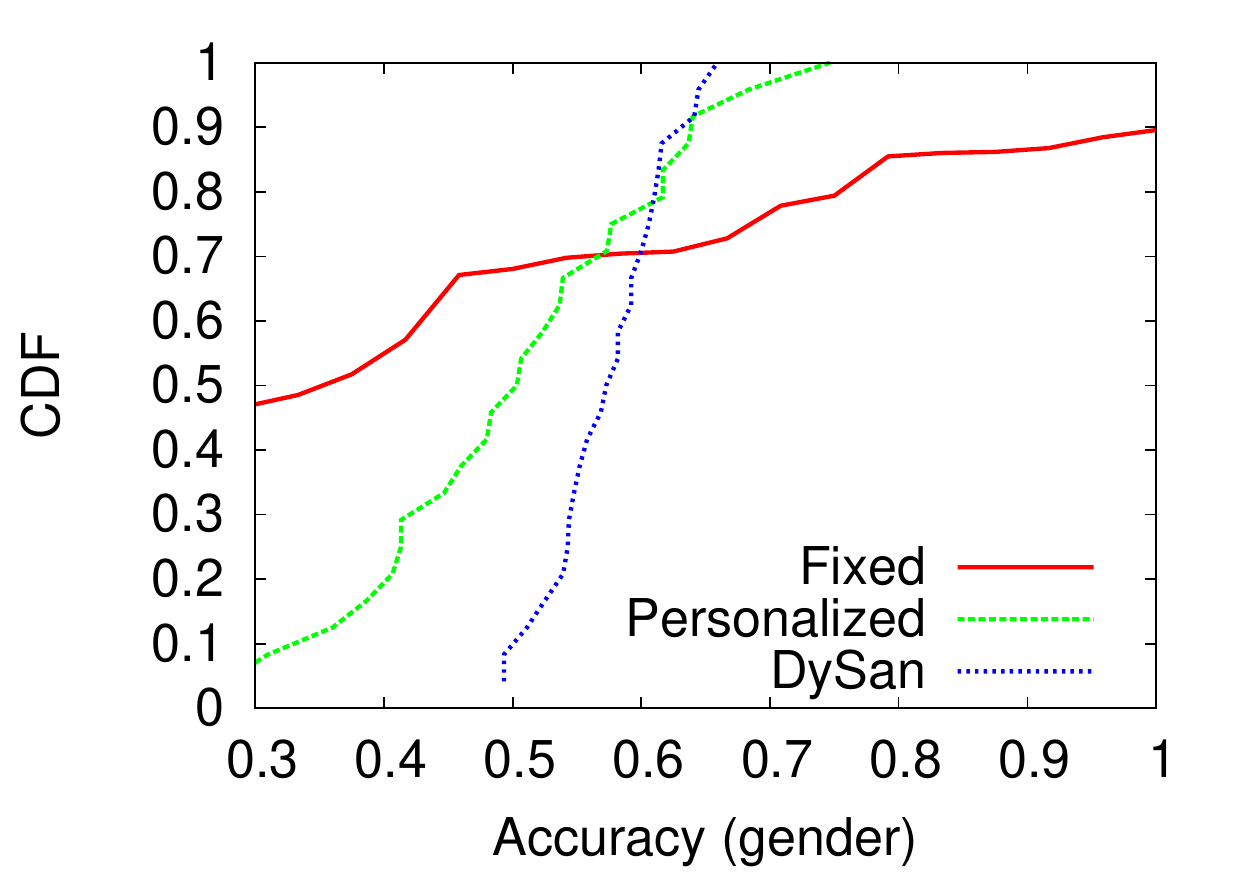}
		\label{fig:inference2MS}
	} 
	\subfloat[MobiAct]{
		\includegraphics[width=0.5\linewidth]{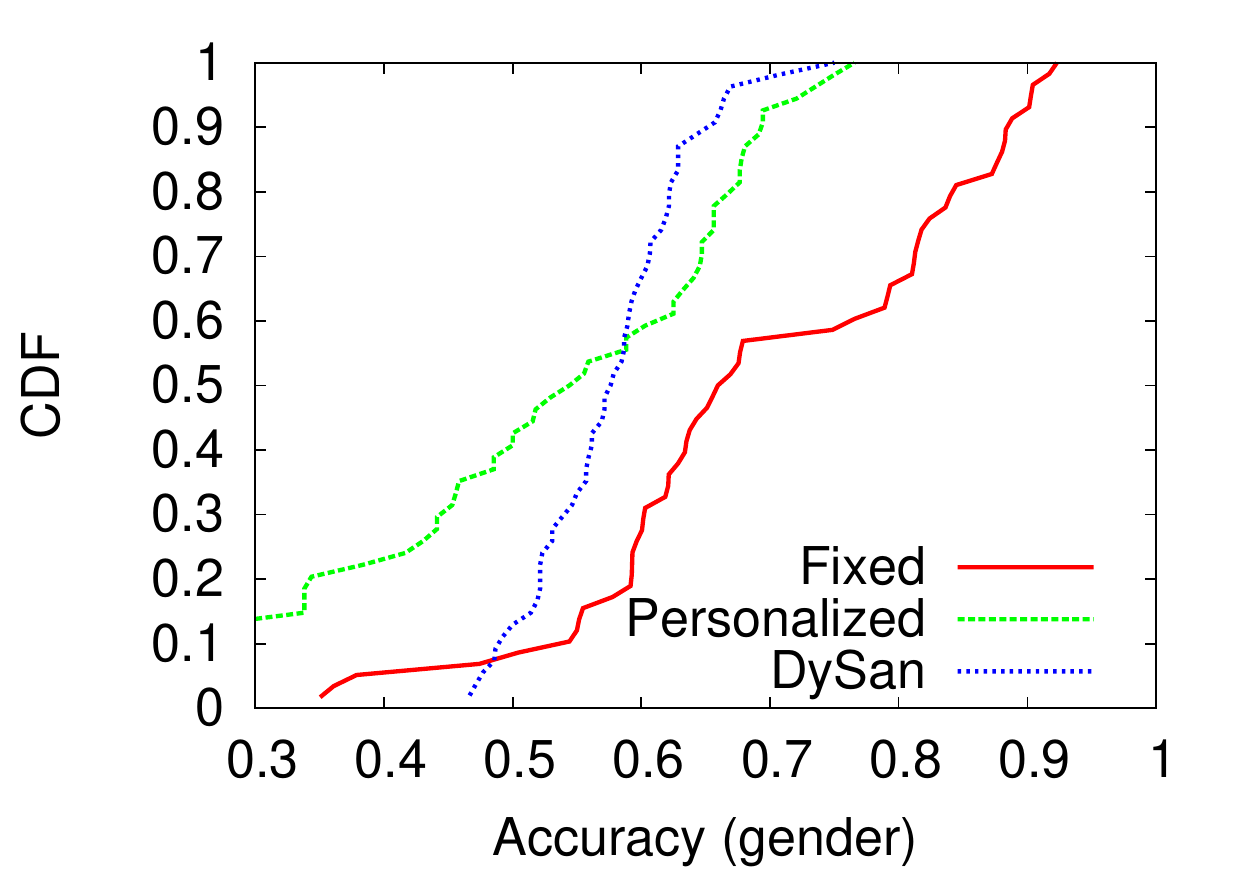}
		\label{fig:inference2MA}
	}
	\vspace{-3mm}
\caption{\small By dynamically adapting the sanitizing model for each user according to the incoming data, \name greatly improved the protection against gender inference (the distribution of the gender accuracy is more centered around 0.5 which corresponds to a random guess).}
\label{fig:cdf-gender}
%\vspace{-2mm}
\end{figure}

We compare these both static solutions against \name where the considered sanitizer model for each user changes according to the incoming data in order to maximize the utility and privacy trade-off over time. 
%\textbf{The selection of the model on the batch of incoming data will be made with the following function we want to maximise:\\}
%
%$$ S(P,U) = xU + yP$$
%\textbf{with  x and y (weight coefficients) $\in$ [0,1], P privacy and U activity accuracy.
%}
%\[
%   P = 1 - |c - p|
%\]
%\textbf{with p gender accuracy, and c ideal gender accuracy, in our case $c=0.5$. The closer the gender accuracy p is to 0.5, corresponding to a random choice between the two genders, the higher the value P is. The two weight coefficients x and y can let the user giving more or less importance on Utility and Privacy. For the next results, we set the values at $x=0.1$ and $y = 0.9$. Figure \ref{fig:selecmotion} \ref{fig:selecmobi} and  in Appendice gives the evolution of the trade-off between Utility and Privacy following the parameter x and y.
%}
%
Figures~\ref{fig:cdf-activity} and~\ref{fig:cdf-gender} depict for both datasets the cumulative distribution (\textit{i.e.}, CDF) of the accuracy of the activity recognition and the gender inference respectively, 
when a fixed, a personalized, and a dynamic sanitizing model is considered.
Firstly, results show that the accuracy in both classification tasks is highly heterogeneous over the population of users. This high heterogeneity reflects the fact that a static model is not well adapted for all users or for all activity performed by the user which motivates our dynamic approach.

Specifically, results show that dynamically adapting the sanitizing model significantly improves the activity recognition compared to using a static model in case of transfer learning (\textit{i.e.}, MobiAct dataset, Figure~\ref{fig:inference1MA}). For MotionSense dataset (Figure~\ref{fig:inference1MS}), most users benefit from an important accuracy with a static model fixed for all users. This result can be explained by the fact that the sanitizing models have been learned with the same users, leading to a learning of the motion characteristics of all the considered users.

For the gender inference, the objective of the sanitizer is to provide an accuracy around 0.5 which corresponds to a random guess for all users.
However, results depicted in Figure~\ref{fig:cdf-gender} clearly shows that a fixed model for all users fails to protect against gender inference. Indeed, the distribution reports a wide range of accuracy over the users where it is possible to infer the gender with 80\% of confidence for 60\% and 20\% of the users for MobiAct and MotionSense dataset, respectively. 
Adopting a personalized sanitizer model for each user decreases the accuracy of the gender prediction compared to a fixed model for all users but the distribution of the accuracy is still large (from 0.3 to 0.75 for MotionSense and from 0.3 to 0.8 for MobiAct).
By dynamically adapting the sanitizing model according to the incoming data, \name greatly improves the protection against gender inference compared to using a fixed model with a sharper distribution centered around 0.5.

%reduces the gender inference (around 20\% and 5\% less accurate in average for MotionSense and MobiAct, respectively).

%(except for the utility on MotionSense dataset where most of the users benefit from an important accuracy which can be explained by the fact that the sanitizing models have been learned with the same users).
%This high heterogeneity reflects the fact that the selected models is not adapted for all users which motivates our personalization approach.
%Specifically, results show that adopting personalized sanitizing model significantly reduces the gender inference (around 20\% and 5\% less accurate in average for MotionSense and MobiAct, respectively).
%In term of utility, accuracy in both cases is instead similar.

%\begin{figure}[t]
%	\centering
%	\subfloat[MotionSense]{
%		\includegraphics[width=0.5\linewidth]{plots/plot_pets_motion/threeplot_dist_motion2.eps}
%		\label{fig:distMotion}
%	} 
%	\subfloat[MobiAct]{
%		\includegraphics[width=0.5\linewidth]{plots/plot_pets_mobi/threeplot_dist_mobi2.eps}
%		\label{fig:disMobi}
%	}
%\caption{\name provides a large variability in terms of distance over all users highlighting the necessity to provide a variety of models to adapt the sanitization.}
%\label{fig:cdf-distance}
%\vspace{-4mm}
%\end{figure}

These results also show the capacity of \name to transfer a learning performed on MotionSense to MobiAct (an activity recognition accuracy around 92\% on average for a gender accuracy around 57\%).
For comparison, we evaluated the transfer learning of GEN using the original sanitizing model learned on MotionSense (and publicly available) to MobiAct dataset. In this case of transfer learning GEN provides an accuracy in term of activity recognition and gender detection around 43\% and 56\%, respectively.
This result shows the limited capacity of GEN to transfer learning from MotionSense to another dataset assuming an over adaptation of the underlying neural network and parameters to the considered dataset.

To complete this analysis, we evaluate the variation of the sanitizing model selection of \name compared to static approaches as well as the number of different models used by \name for each user in Appendix~\ref{sec:dynamicSelection}. We also quantify the possibility to use the set of selected models as a fingerprint to identify each user in Appendix~\ref{subsec:leakage}.

\begin{table*}
\begin{minipage}{\linewidth}
      \centering
      \begin{minipage}{0.45\linewidth}
          \begin{figure}[H]
\includegraphics[scale=0.46,trim=0 0 0 0]{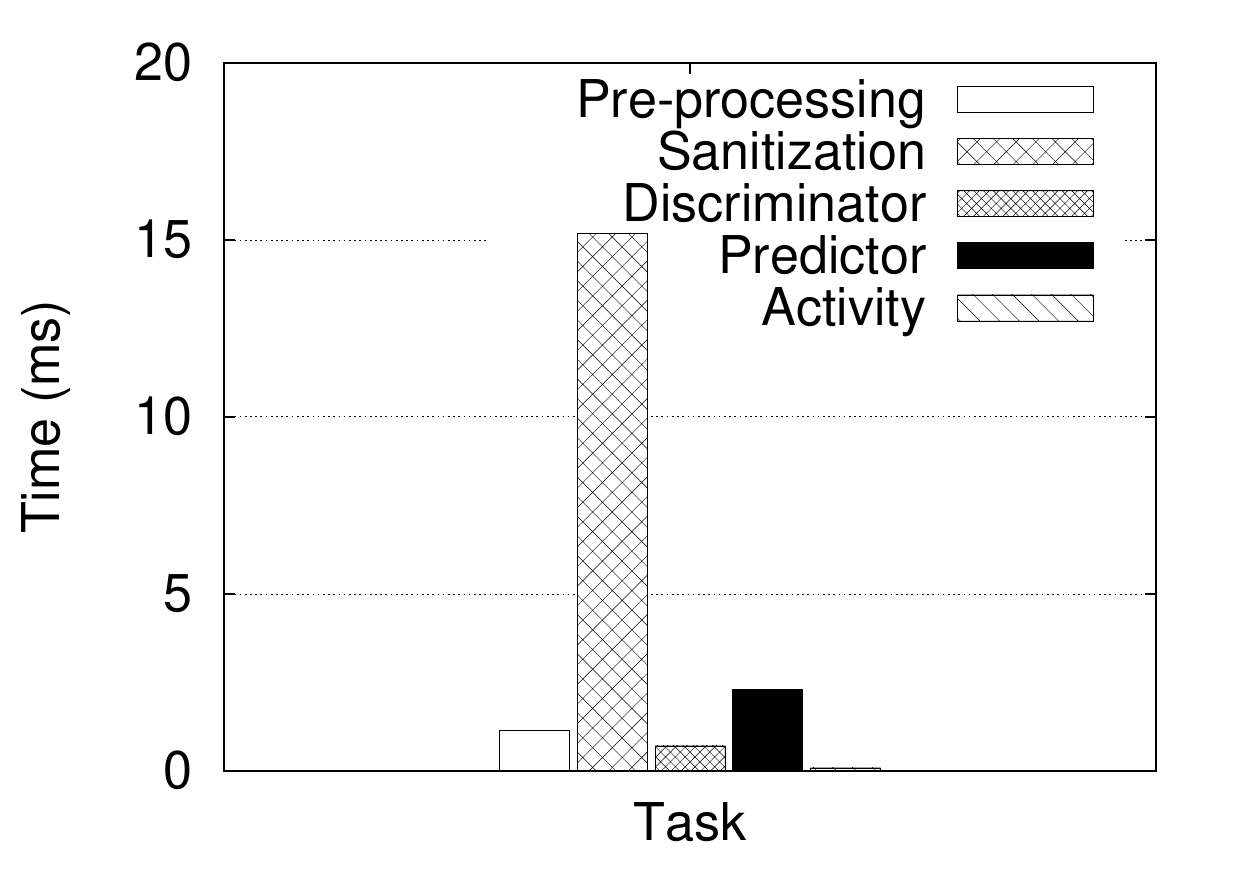}
\caption{\small The limited cpu overhead of the sanitation of \name is compatible to real-time processing on smartphone.}
\label{fig:cpu_time}
\end{figure}
      \end{minipage}
      \hspace{4mm}
      \begin{minipage}{0.45\linewidth}
          \begin{figure}[H]
\includegraphics[scale=0.46,trim=0 0 0 0]{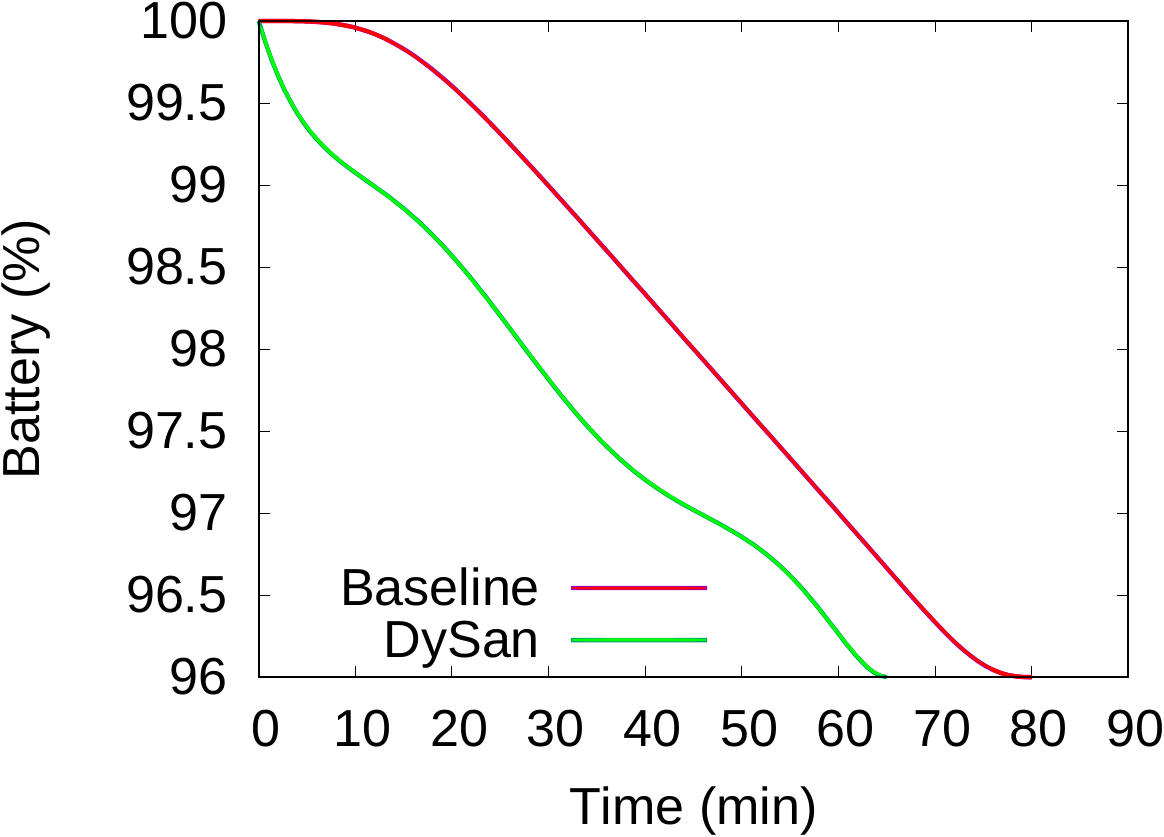}
\caption{\small The impact of \name on energy consumption is limited (1\% less battery after 1 hour).}
\label{fig:energy}
\end{figure}
      \end{minipage}
  \end{minipage}
  %\vspace{-2mm}
\end{table*}

%\begin{figure}[t]
%	\centering
%	\includegraphics[width=7.5cm]{plots/tasks.eps}
%	\caption{The limited cpu overhead of the sanitation of \name is compatible to real-time processing on smartphone.}
%	\label{fig:cpu_time}
%\vspace{-4mm}
%\end{figure}

\subsection{Performance as measured on devices} 
\label{sec:cost}

We now evaluate the cost of operating \name on a smartphone.
\name protects the sensitive attribute while ensuring an accurate activity recognition and minimal data distortion.
However, applying the sanitizing at run time on the mobile introduces an overhead.
We do not consider the overhead of the learning as it is a one time operation.
\name evaluates multiple sanitizing models (i.e., according to each $\alpha$ and $\lambda$ hyperparameter explored) before selecting the one that produces the best compromise between utility and privacy.
Consequently, the overhead associated with the sanitizing of raw data depends on the number of considered models.

Figure~\ref{fig:cpu_time} describes the time (ms) spent by a Xiaomi Redmi Note 7 (equipped with a Qualcom Snapdragon 660 and 3 GB of memory running a java application using Pytorch 1.6) on each task associated to a single sanitizing model of a window of incoming data (i.e., 2.5 seconds of data). Specifically, these tasks include the pre-processing of signal, the sanitizing of raw data, the evaluation of the privacy and the utility on the sanitized data respectively by the discriminator and the predictor, and the classification of the activity performed by the user from the raw data.
Excepting the pre-processing which is performed only once for a window of data, the other tasks have to be repeated for each explored sanitizing model.
Results show that applying a sanitizing model once spends most of the time while all operations require 19 ms.
Considering 20 or 36 sanitizing models increases this time to 366 ms and 658 ms, respectively.
Although this processing is compatible with real-time processing (i.e., data processed after each data window), the number of models explored should be chosen to limit the overload.

We also evaluate the impact of running \name on the energy consumption on the smartphone.
Figure~\ref{fig:energy} reports the decrease in the battery charge over time for a baseline where no operation is performed on the smartphone, and for a real-time processing of \name (i.e., after each window of raw data, and exploring 36 sanitizing models before to select the best one). In both cases, the screen remained on during the experiment.
Results show that \name consumed 1\% more battery after 1 hour, which stays a reasonable energy consumption.

\section{Related Works}
\label{sec:relatedworks} 
With the availability of wearable and personal devices, there have been a growing research on the use of collected data for quantifying various aspects of personal life, such as the number of calories consumed, the blood pressure, etc.
 A growing literature concern the use of data for predicting the physical activities performed by users, let it be for either medical, insurance or various other reasons. We refer the interested reader to the surveys
 \cite{nweke2018deep} and \cite{ramasamy2018recent} on machine learning and deep learning techniques applied for predicting the type of activities performed.

In this section, we compare our approach with other existing techniques that protect sensitive information in sensor data while retaining data utility. Our approach is closely related to Gansan \cite{aivodji2019agnostic}, however, our framework goes beyond, by considering the data utility with respect to the Predictor network, in addition to the application on motion sensor data.
Next, we focus on approaches used as baselines previously in the paper. \cite{Malekzadeh2018} is the only one which focus on the gender as the sensitive information. \cite{jourdan2018toward} and \cite{Malekzadeh_2019} in their case, focus on the re-identification only while \cite{olympus_raval} apply their approach on several applications like object recognition or action recognition with several data types such as images or motion sensors. In the case adversarial approach that use autoencoders, the sensitive information can be extracted from the representation produced by the encoder \cite{liu2017deeprotect}, the decoder \cite{olympus_raval} which also correspond to our approach, or both the encoder and the decoder \cite{Malekzadeh_2019} for data sanitization. 
Specific to the sensor data generation, SenseGen \cite{alzantot2017sensegen} is a deep learning architecture for protecting users privacy by generating synthetic sensor data. Unfortunately, they did not provide any guarantee on the protection. 

To enlarge with other applications protecting sensitive informations using adversarial methods, \cite{chen2018vganbased} use a VGAN to transform face images in order to hide facial expression of the users that can be used to reveal their identity while preserving generic expressions. Adversarial approaches can also be used to hide sensitive information such as text in images \cite{edwards2015censoring} or identity information in the fingerprints \cite{oleszkiewicz2018siamese}.

From a broader privacy perspective, \cite{tripathy2019privacy} proposes an adversarial network technique to minimize the amount of mutual information between a
 sensitive attribute and useful data while bounding the amount of distortion introduced. They applied their solution on a synthetic and a computer vision
 dataset. Inspired from \cite{tripathy2019privacy}, authors in \cite{romanelli2019generating} have developed a method for learning an optimal privacy
 protection mechanism also inspired from GAN, which they have applied to location privacy. In \cite{park2018data}, authors have proposed an approach called table-GAN, which aim at preserving privacy by generating synthetic data. By suppressing \textit{one-to-one} relationship and limiting the quality of dataset reconstruction re-identification attacks are rendered less performant.
 They compared their approach with standard privacy techniques such as k-anonymity t-cl and closeness.
 
Apart from techniques using adversarial approach to protect sensitive information on sensor data, \cite{menasria} proposes two privacy preserving mechanisms based on clustering algorithms called Hierarchical Agglomerative Clustering to compress amount of disclosed data so that the amount of sensitive information can be reduced. \cite{zarepour} in their case, develop a framework for images data made on wearable cameras that can protect sensitive information such as face, objects or locations thanks to a neural network that detects the sensitive objects which will then be blurred or deleted. Rather than focusing on re-idenfication, \cite{chakraborty} investigate what data to share, in such a way that certain kinds of inferences cannot be down. They propose \textit{ipShield} that obfuscate data according to the quantification of an adversary's knowledge regarding a sensitive inference.

\section{Conclusion}
\label{sec:conclusion} 

We presented \name, a privacy-preserving framework which 
sanitizes motion sensor data in order to prevent unwanted 
inference of sensitive information.
At the same time, \name preserves as much as possible the 
useful information for activity recognition and other 
estimators of physical activity monitoring. 
Results show that \name drastically reduces the risk of gender inference without impacting the ability to detect the activity or to monitor the number of steps.
We also show that the dynamic sanitizing model selection of \name successfully adapts the protection to each user over time according to the evolution of the incoming data.
Moreover, we show that the overhead introduces on the smartphone to sanitize the data is compatible with real-time processing while keeping a reasonable energy consumption.
Lastly, we compared our approach with existing 
approaches and demonstrated that \name provides better control over privacy-utility trade-off.

We investigated the possibility to extend \name to take into account multiple sensitive attributes. Our preliminary results by adding several discriminators accounted in the loss function of the sanitizer's training are encouraging, however, we are limited by the small size of the available datasets. Indeed, making the sanitizing models more complex requires more data to capture the specificity of each use case.

\bibliographystyle{norunsrt}
\bibliography{bib}

%\section*{Declarations}

%\begin{itemize}
%    \item \textbf{Funding} The authors did not receive support from any organization for the submitted work.\\
%\item \textbf{Conflicts of interest} The authors declare that they have no conflict of interest.\\
%\item \textbf{Availability of data and material} The datasets used are publicly available.\\
%\item \textbf{Code availability} The code is available at \url{https://github.com/DynamicSanitizer/DySan}
%\end{itemize}

%\input{bib_ecml.tex}

\newpage 

%\clearpage
\section*{Appendices}
\appendix

\section{Neural Network Architecture}
\label{sec:nnarchi}

We provide in this section details about the underlying neural networks of \name.

\subsection{Discriminator Net}
\label{sec:dnetarchi}

\begin{footnotesize}
{\begin{enumerate}
\setlength{\itemsep}{0pt}%
\setlength{\parskip}{0pt}%
\item Input (125,6)
\item Conv1D (64, kernel\_size=6, stride=1, activation=ReLU)
\item AvgPool1D(kernel\_size=2, stride=2)
\item BatchNorm1D(100, eps=1e-05, momentum=0.1)
\item Dropout(p=0.5)
\item Dense(64, activation=ReLU)
\item Dense(2, activation=softmax)
\end{enumerate}}
\end{footnotesize}

\subsection{Predictor Net}
\label{sec:pnetarchi}

\begin{footnotesize}
{\begin{enumerate}
\setlength{\itemsep}{0pt}%
\setlength{\parskip}{0pt}%
\item Input (125,6)
\item Conv1D (100, kernel\_size=6, stride=1, activation=ReLU)
\item AvgPool1D(kernel\_size=2, stride=2)
\item BatchNorm1D(100, eps=1e-05, momentum=0.1)
\item Conv1D(100, kernel\_size=5, stride=1, activation=ReLU)
\item AvgPool1d(kernel\_size=2, stride=2)
\item Conv1D(160, kernel\_size=5, stride=1, activation=ReLU)
\item AvgPool1d(kernel\_size=2, stride=2)
\item Conv1D(160, kernel\_size=5, stride=1, activation=ReLU)
\item AvgPool1d(kernel\_size=2, stride=2)
\item Dropout(p=0.5)
\item Dense(64, activation=ReLU)
\item Dense(4, activation=softmax)
\end{enumerate}}
\end{footnotesize}

\subsection{Sanitizer Net}
\label{sec:snetarchi}

\begin{footnotesize}
{\begin{enumerate}
\setlength{\itemsep}{0pt}%
\setlength{\parskip}{0pt}%
\item Input (125,6)
\item Conv1D (64, kernel\_size=6, stride=1,)
\item Conv1D (128, kernel\_size=5, stride=1)
\item Dense(128)
\item Dense(64, activation=LeakyReLU(0.01))
\item Dense(64)
\item Dense(128)
\item Deconv1D (128, kernel\_size=5, stride=1)
\item Deconv1D (64, kernel\_size=5, stride=1, activation=softmax)
\end{enumerate}}
\end{footnotesize}

\newpage

\section{Sanitized Data Distortion}
\label{sec:deformation}

Table \ref{tab:other-metrics} gives complementary results concerning the similarity analysis of the data sanitized between the different baselines, with simple quantitative measures. Here the raw measures plus the percentage relative error are given for each baselines. Even if those metrics gives few information about the shapes of the signals, we can still observe that Olympus, the only baselines that does not take into account the distortion of the data during training, is the one that have his measures very far from the raw data. For example the standard deviation is almost fives times higher than the original data showing a very noisy signal.

\begin{figure*}
\begin{table}[H]
\begin{tabular}{|c|c|c|c|c|c|}
  \hline
   & Mean & Std & Skewness & Kurtosis & Energy  \\
  \hline
  Raw & 0.81 & 0.47 & 1.65 & 4.81 & 139.06  \\
    \hline
  DySan & 0.68 (-15.9\%) & 0.77 (+62.9\%) & 0.40 (-75.7\%) & 1.28 (-73.5\%) & 230.87 (+66.0\%)\\

    \hline
  GEN & 0.28 (-65.4\%) & 0.12 (-74.7\%) & 0.51 (-69.2\%) & 0.08 (-98.3\%) & 12.11 (-91.3\%) \\
    \hline
  Olympus & 5.40 (+566.4\%) & 2.52 (+433.1\%) & 0.61 (-62.8\%) & 0.29 (-94.0\%) & 4631.47 (+3230.5\%) \\
   \hline
  MSDA & 0.54 (-33.5\%) & 0.24 (-49.9\%) & 0.41 (-75.2\%) & -0.11 (-102.2\%) & 51.87 (-62.7\%) \\
  \hline
\end{tabular}
\end{table}
\vspace{-4mm}
\caption{\small Similarities metric between the raw data and the different baselines. Mean, standard deviation (std), skewness, kurtosis, energy are given in percentage of relative error.}
\label{tab:other-metrics}
\end{figure*}

\section{Heterogeneous Activity Classification}
\label{sec:classificationDetails}

The accuracy of the classification is not uniform for all activities.
Table~\ref{tab:activitytable} details the True Positives and False Positives of this classification for \name on MotionSense dataset.
This table also reports the percentage of data in the dataset for each activity.
We observe that the accuracy of the classification depends on the performed activity. This heterogeneity is a direct result of the unbalanced classes. Specifically, the walking activity has the highest precision which corresponds to the activity with the largest amount of data, while other activities contains less data and depicted lower good predictions.
This difference in terms of good prediction between walking and other activities can also be explained by a calibration of the size window adapted for the walk (see Section \ref{sec:sysmodel}).

\begin{table}[H]
\centering

\begin{tabular}{|c|c|c|c|c|}
  \hline
   & TP & FP & Precision & Data percentage \\
  \hline
  Downstairs & 221 & 112 & 66.4 & 17.2  \\
    \hline
  Upstairs & 223 & 198 & 53.0 & 20.5  \\
    \hline
  Walking & 918 & 74 & 92.5 & 44.9  \\
    \hline
  Jogging & 216 & 212 & 50.5 & 17.4  \\
    \hline
\end{tabular}
\caption{\small True Positive, False Positive, Precision and percentage of data for each activity of Dysan (MotionSense dataset).}
\label{tab:activitytable}
\end{table}

%\begin{table}[H]
%\centering
%\begin{tabular}{|c|c|c|c|c|}
%  \hline
%   & TP & TN & FP & FN \\
%  \hline
%  Raw & 1235 & 917 & 17 & 4  \\
%    \hline
%  DySan & 694 & 522 & 402 & 528  \\
%    \hline
%  GEN & 1234 & 883 & 51 & 5  \\
%    \hline
%  Olympus & 825 & 719 & 215 & 414  \\
%    \hline
%  MSDA & 771 & 697 & 237 & 468   \\
%  \hline
%\end{tabular}
%\caption{True Positive, True Negative, False Positive, False Negative values for gender classification of every baselines}
%\label{tab:detailgendertable}
%\end{table}

\section{Dynamic sanitizing mode selection}
\label{sec:dynamicSelection}

\begin{figure}[!h]
	\centering
	\subfloat[MotionSense]{
		\includegraphics[width=0.5\linewidth]{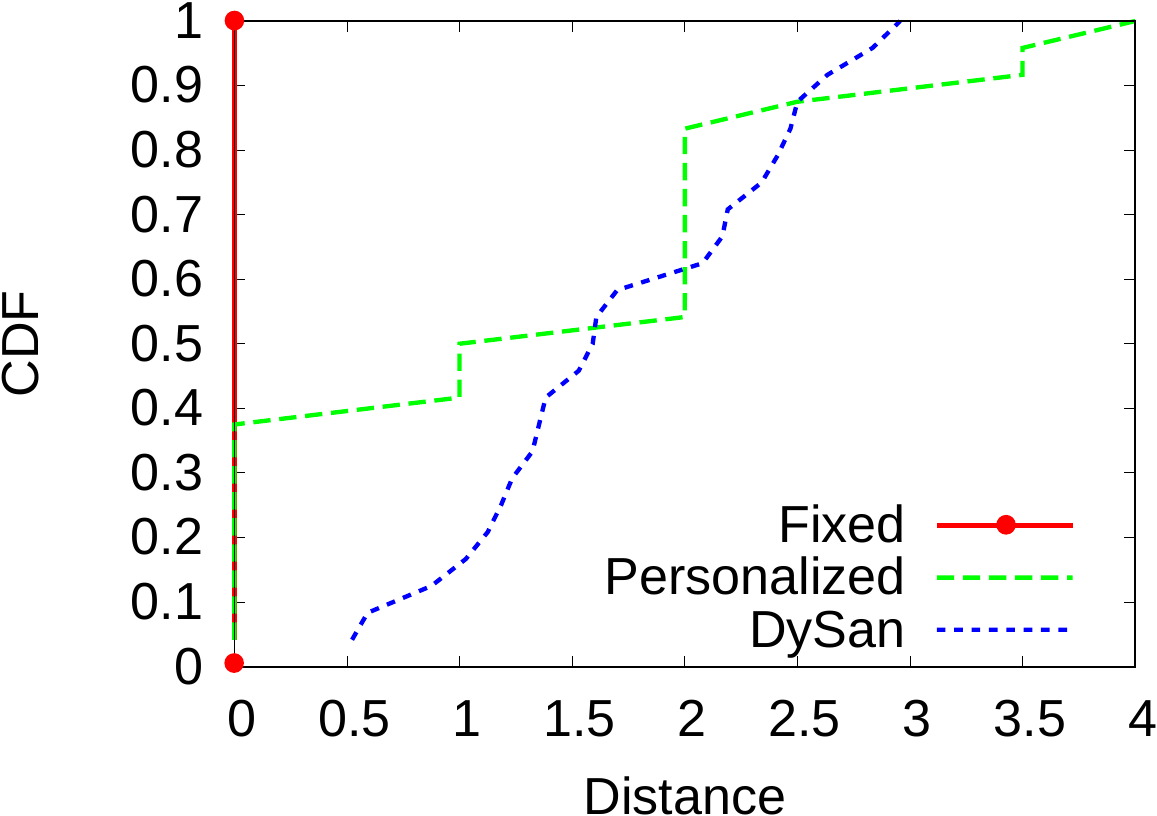}
		\label{fig:distMotion}
	} 
	\subfloat[MobiAct]{
		\includegraphics[width=0.5\linewidth]{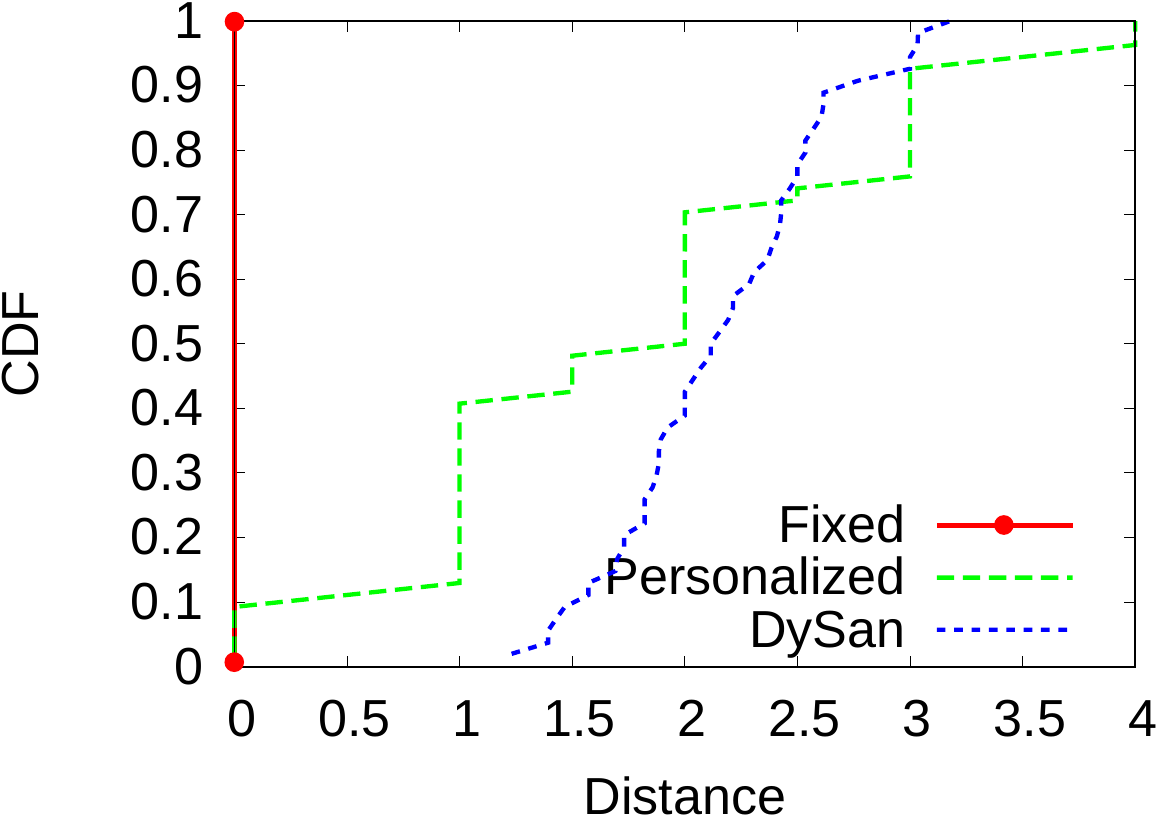}
		\label{fig:disMobi}
	}
\caption{\small \name provides a large variability in terms of distance over all users highlighting the necessity to provide a variety of models to adapt the sanitization.}
\label{fig:cdf-distance}
%\vspace{-4mm}
\end{figure}

We evaluate the variation of the sanitizer model selection of \name compared to static approaches using either one model fixed for all users or one personalized model for each user. To achieved that, we measure the distance between the hyperparameters $\alpha$ and $\lambda$ corresponding to the best privacy and utility trade-off on average for all users (\textit{i.e.}, the model fixed for all users) and the model selected for each user (\textit{i.e.}, a personalized model) or according to the incoming data (\textit{i.e.}, the model dynamically selected by \name). Figure~\ref{fig:cdf-distance} reports the distribution of this distance for both datasets. 
Results show that almost 40$\%$ of the users of MotionSense dataset have a personalized sanitized model which corresponds to the model providing the best trade-off on average for all users. In addition, for both datasets, results show a large variability in term of distance over all users highlighting the necessity to provide a variety of models to adapt the sanitization.

\begin{figure}[t]
	\centering
	\includegraphics[width=6.5cm]{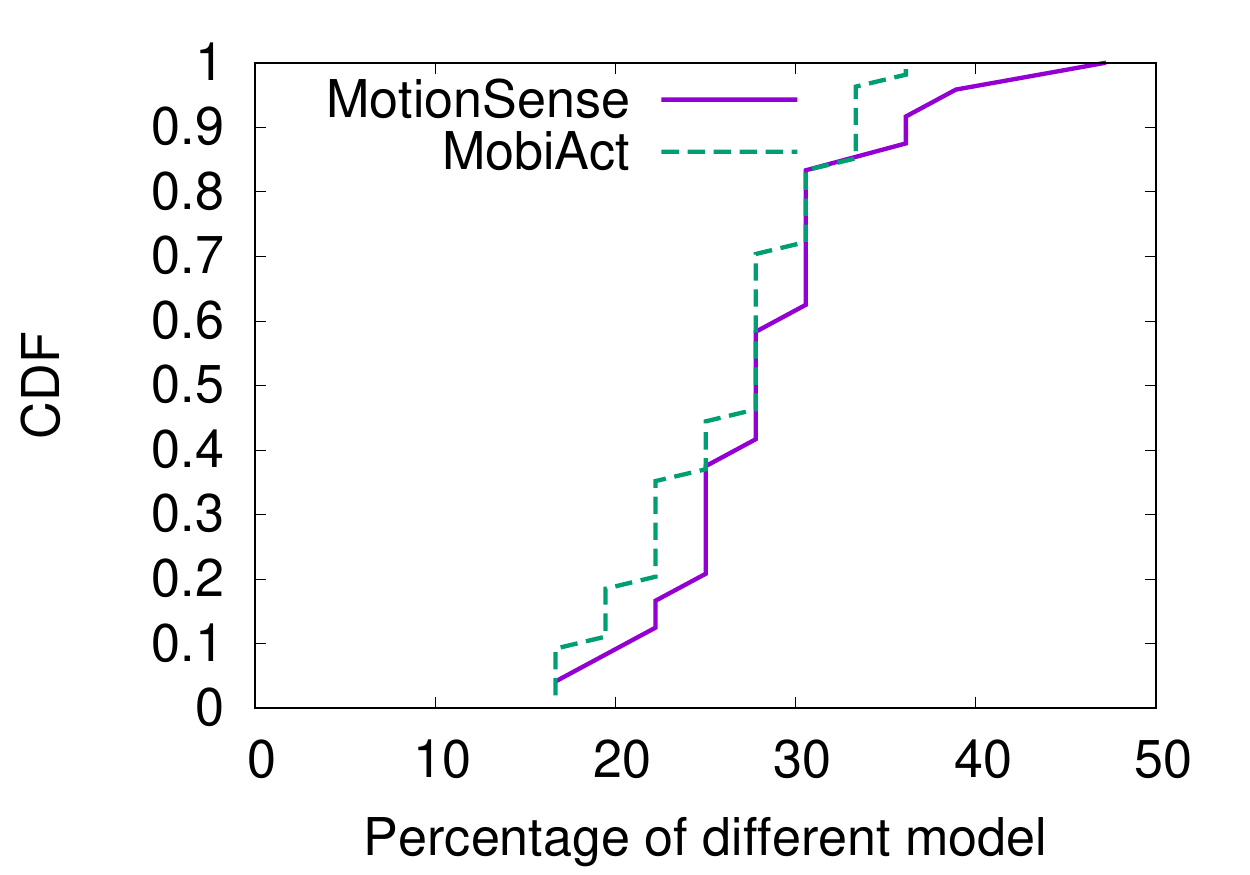}
	\caption{\small The data of each user is sanitized with a wide variety of models (from 20\% to 50\% of all the models) showing that \name successfully adapts the sanitization according to the evolution of the incoming data.}
	\label{fig:count_model}
%\vspace{-4mm}
\end{figure}

To complete this analysis, we also counted the number of different models used by \name for each user. Figure~\ref{fig:count_model} depicted for both datasets the distribution of the percentage of all possible sanitized models (36 in our experiment as presented Section~\ref{sec:methodo}) selected by \name for each user. Results show a large range of number of different models selected ranging from 20\% to 50\%. This result show that \name successfully adapts the sanitization according to the evolution of the incoming data.

%\section{Model selection for \name}
\section{Utility and privacy trade-off selection for \name}
\label{sec:tradeoff}

\begin{figure}[H]
	\centering
	\subfloat[MotionSense]{
		\includegraphics[width=0.5\linewidth]{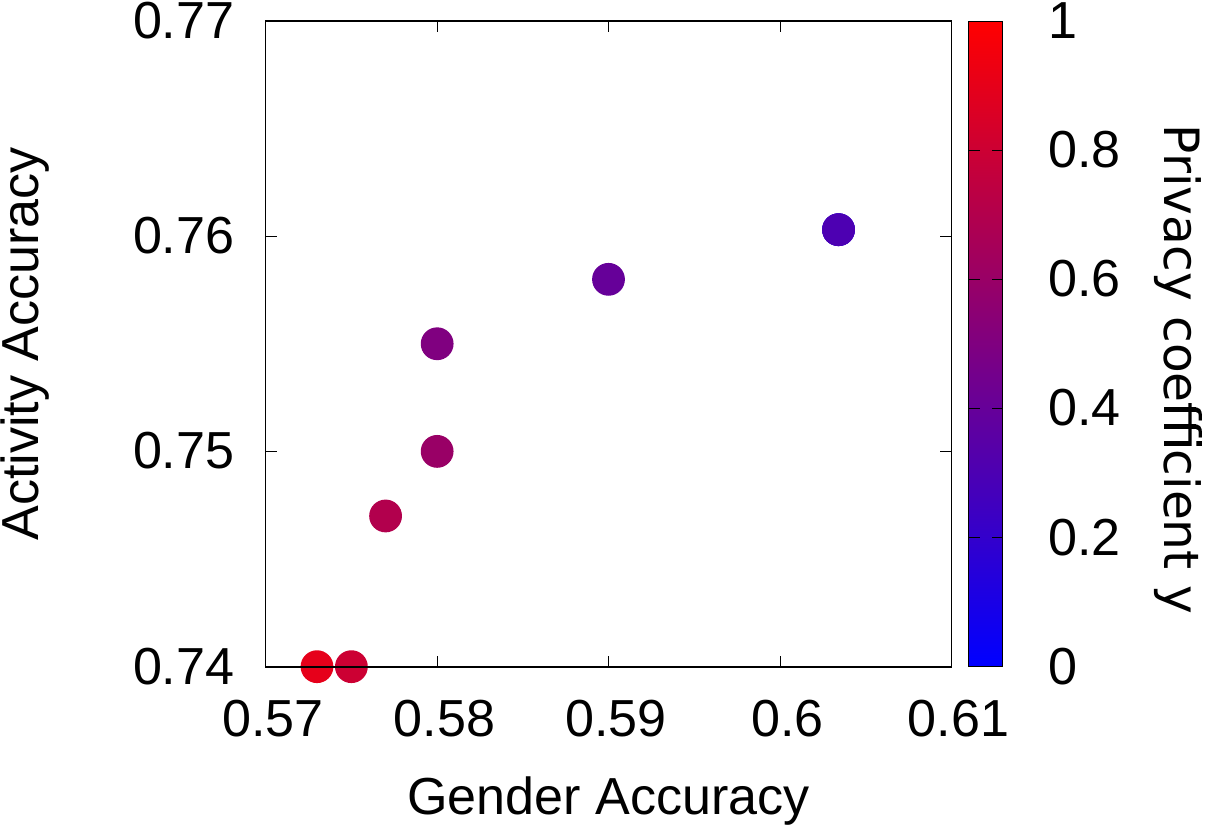}
		\label{fig:selecmotion}
	} 
	\subfloat[MobiAct]{
		\includegraphics[width=0.5\linewidth]{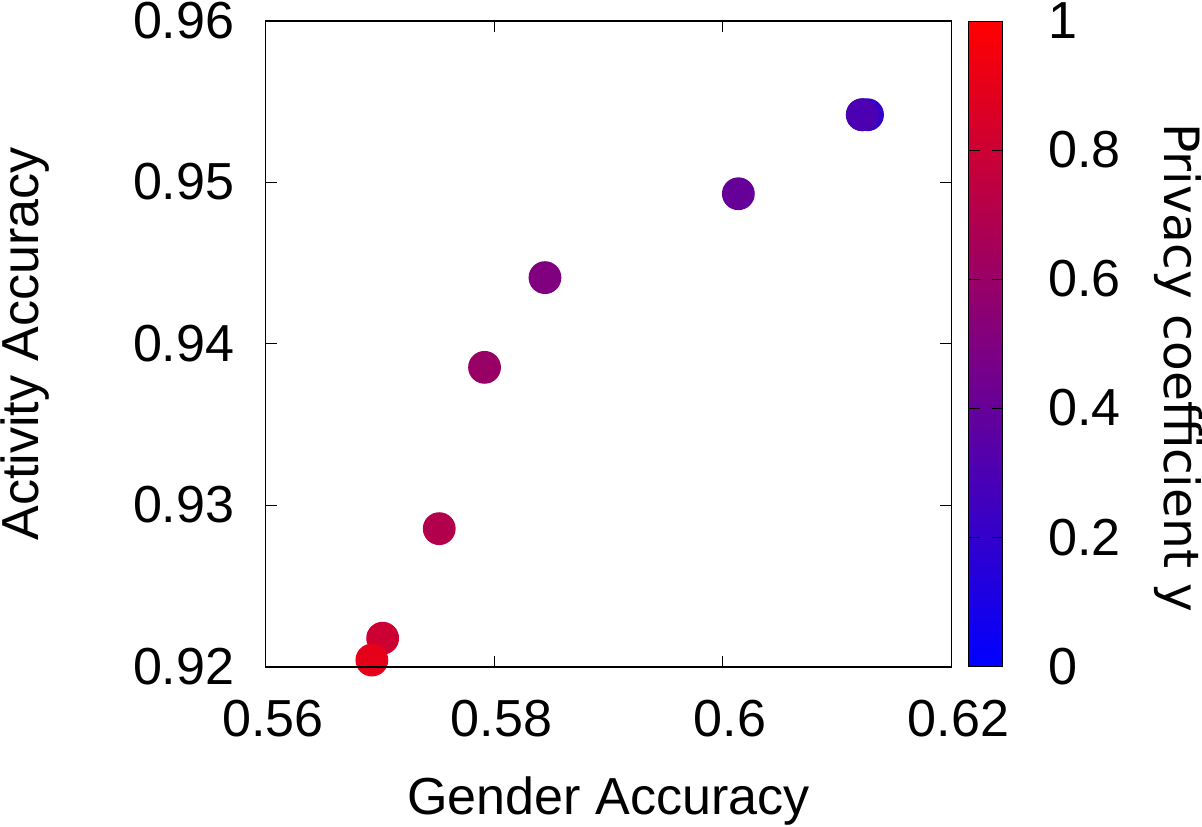}
		\label{fig:selecmobi}
	}
\caption{\small The variation of the Privacy coefficient y from 0.1 to 0.9 implies a variation of the trade-off between Utility and Privacy. For both dataset, when y increase, the Privacy increase (the gender accuracy decrease) and the Utility decrease (the activity accuracy decrease)}
\label{fig:selection-model}
%\vspace{-4mm}
\end{figure}

As described in Section~\ref{sec:online}, the best sanitizer model is selected according to the definition of the utility and privacy trade-off defined by weight coefficients x and y. Figure~\ref{fig:selection-model} depicts the evolution of the utility and privacy trade-off according to x and y for both datasets. 
%\textbf{The way of maximizing the trade-off between utility and privacy implies that each user tends to remain close to his optimal trade-off even with a large variation of x and y. It implies a low average variability trade-off even if the framework with a large set of different models can allow a large variability.}

\section{Information leakage in model selection}
\label{subsec:leakage}

\begin{figure}[t]
	\centering
	\includegraphics[width=6.5cm]{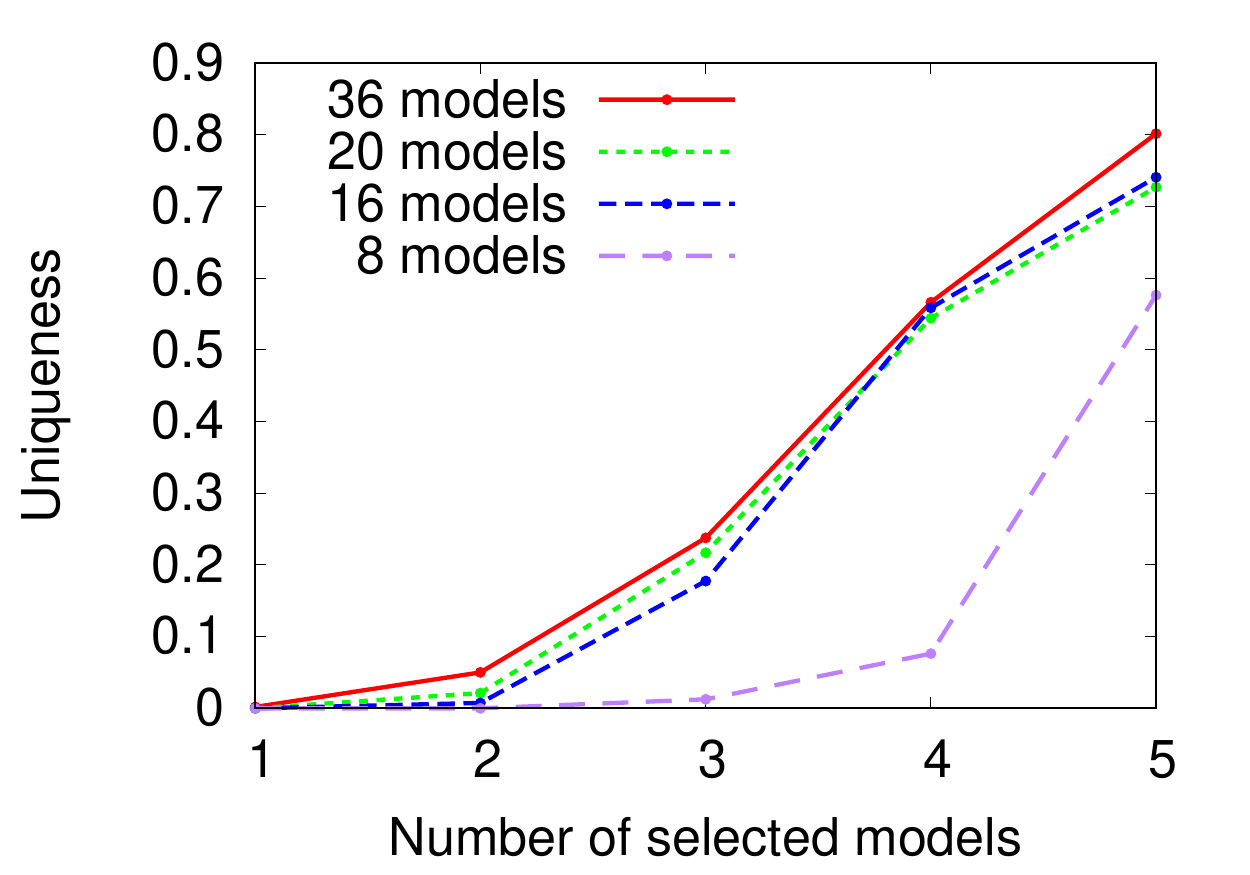}
	\caption{\small The uniqueness of the selected models remains low  for fingerprints with less than 5 models, and depends on the number of available sanitizing models for the selection.}
	\label{fig:uniqueness}
%\vspace{-4mm}
\end{figure}

As \name dynamically selects the sanitizing model to use for each window of incoming data, the set of selected models could be leveraged to identify each user. 
Indeed, this set of sanitizing models chosen by a user could act as a unique fingerprint. 
To evaluate this potential information leakage, we quantify the uniqueness following the methodology presented in~\cite{demontjoye2013unique}.
More precisely, the uniqueness for each user is estimated as the percentage of 100 random sets of $p$ selected sanitizing models that are unique.
Figure~\ref{fig:uniqueness} reports for MobiAct dataset the distribution of the uniqueness with $p$ (\textit{i.e.}, the size of fingerprint) from 1 to 5 and with different number of sanitizing models available for the selection.
As expected, results show that the larger the fingerprint, the more unique the behaviour of a user becomes. However, at least 5 models are needed to have a strong confidence (around 80\% of uniqueness) when 36 sanitizing models are exploited.
To reduce this uniqueness, a lower number of sanitizing models (i.e., through the hyperparameters values explored in the training phase) should be proposes. Indeed, less choice for model selection leads to have more users who share common models. 
Results show that exploiting less available sanitizing models reduces the uniquenes.
%Seb: j'ai mis cette phrase en commentaire car le contexte est très différent ici comme le nombre de modèles est beaucoup plus petit que le nombre de localisations dans les travaux de Montjoye
%This probability of uniqueness is less important than the one observed in other data such as location~\cite{}.

Reducing the number of sanitizing models by covering less hyperparameter values limits the achievable space for the utility and privacy trade-off. Consequently, a degradation of the accuracy for both the activity detection and the gender interference is observed. Table \ref{tab:accuracyselection} presents the performances obtained with different number of sanitizing models available for the selection. Results show that from 36 to 20 sanitizing models, the accuracy in activity recognition decreases by only 3\% and increase by 2\% the gender inference.

%\begin{figure}[h]
%	\centering
%	\includegraphics[width=7.5cm]{plots/pets_v2/cdf_modelsuser_mobi.eps}
%	\caption{For MobiAct dataset when all models are used, only one user select a model that no one use. Two models are not used by any users}
%	\label{fig:modelselection1}
%\vspace{-4mm}
%\end{figure}

%\textbf{Antoine: still finishing this section}
%As a possible countermeasure, we have reduced the number of models available for each user in order to increase the number of users per model.
%Seb: j'ai mis en commentaires la phrase suivante qui serait à clarifié
%We have analysed three different configurations with less models available for each users based on the choice of the hyperparameter $\alpha$ and $\lambda$. 
%As expected, covering less hyperparameters values reduce the achievable space for the utility / privacy trade-off. 
%Table \ref{tab:accuracyselection} presents the performances obtained with the decrease of number of models available for each user.
%\textbf{To address this issue, we can reduce the number of models available for each user in order to increase the number of user for each model. Then, we propose three different configurations with less models available for each users based on the choice of the hyperparameter $\alpha$ and $\lambda$ chosen. Covering less hyperparameters values has the effect of reducing the performances in utility and privacy. Table \ref{tab:accuracyselection} evaluate the performances as the number of models available for each user decrease:}

Information leakage in model selection leading to user re-identification is only possible if the adversary is able to characterize each selected sanitizing model from the sanitized data. In this case, the adversary could maintain a fingerprint per user to conduct its re-identification attack.
To evaluate this capability, we measure the level of distortion using the Dynamic Time Warping of the sanitized data for each sanitizing model. 
Over all sanitizing models, our results show a very low standard deviation of the DTW. This low value indicates a small difference in terms of distortion when different sanitizing models are exploited, thus making it difficult for an adversary to identify the selected model from the sanitized data. This re-identification attack consequently seems difficult to achieve.

\begin{table}[H]
\centering

\begin{tabular}{|c|c|c|}
  \hline
    & Activity accuracy (\%) & Gender accuracy (\%)  \\
  \hline
  36 models & 92 & 57  \\
      \hline
20 models & 89 & 59  \\
    \hline
 16 models & 88 & 63  \\
    \hline
      8 models & 86 & 66  \\
    \hline
\end{tabular}
\caption{\small Reducing the number of sanitizing models available for the selection decreases the accuracy in activity recognition while increasing the accuracy in gender inference.}
\label{tab:accuracyselection}
\end{table}

\end{document}